\documentclass[a4paper,11pt]{article}
\usepackage{jheppub}
\usepackage{bigints}
\usepackage{float} 
\usepackage{paralist}
\usepackage{multirow}
\usepackage{array}
\newcolumntype{P}[1]{>{\centering\arraybackslash}p{#1}}
\usepackage[T1]{fontenc}    
\usepackage{hyperref}       
\usepackage{url}            
\usepackage{booktabs}       
\usepackage{amsfonts}       
\usepackage{nicefrac}       
\usepackage{microtype}      
\usepackage{lipsum}		
\usepackage{graphicx}
\usepackage{natbib}
\usepackage{doi}
\usepackage{graphicx}
\usepackage{caption}
\usepackage{subcaption}
\usepackage{epsf}
\usepackage{bm}
\usepackage{amsmath}
\usepackage{amsfonts}
\usepackage{graphicx}
\usepackage{tabularx}
\usepackage{color}%
\providecommand{\U}[1]{\protect\rule{.1in}{.1in} }

\newcommand{\be}{\begin{equation}}
\newcommand{\ee}{\end{equation}}

\newcommand{\mincir}{\raise
-3.truept\hbox{\rlap{\hbox{$\sim$}}\raise4.truept\hbox{$<$}\ }}
\newcommand{\magcir}{\raise
-3.truept\hbox{\rlap{\hbox{$\sim$}}\raise4.truept\hbox{$>$}\ }}

\providecommand{\U}[1]{\protect\rule{.1in}{.1in}}

\title{Cosmology under the fractional calculus approach: a possible $H_0$ tension resolution?}

\author[a,b]{Genly Leon\footnote{Speaker at Corfu Summer Institute 2022 "School and Workshops on Elementary Particle Physics and Gravity",\\
  28 August - 1 October 2022\\
  Corfu, Greece}}
\author[c]{Miguel A. Garc\'ia-Aspeitia}
\author[c]{Guillermo Fernandez-Anaya}
\author[d]{Alberto  Hern\'andez-Almada}
\author[e]{Juan Maga\~na}
\author[f]{Esteban Gonz\'alez}
\affiliation[a]{Departamento  de  Matem\'aticas,  Universidad Cat\'olica del Norte, Avda.   Angamos  0610,  Casilla  1280  Antofagasta,  Chile}
\affiliation[b]{Institute of System Science, Durban University of Technology, PO Box 1334,
Durban, 4000, South Africa}
\affiliation[c]{Depto. de F\'isica y Matem\'aticas, Universidad Iberoamericana Ciudad de M\'exico, Prolongaci\'on Paseo \\ de la Reforma 880, M\'exico D. F. 01219, M\'exico}
\affiliation[d]{Facultad de Ingenier\'ia, Universidad Aut\'onoma de
Quer\'etaro, Centro Universitario Cerro de las Campanas, 76010, 
Santiago de Quer\'etaro, M\'exico}
\affiliation[e]{Escuela de Ingenier\'ia, Universidad Central de Chile, Avenida Francisco de Aguirre 0405, 171-0164 La Serena, Coquimbo, Chile}
\affiliation[f]{Direcci\'on de Investigaci\'on y Postgrado, Universidad de Aconcagua, Pedro de Villagra 2265, Vitacura, 7630367 Santiago, Chile}
\emailAdd{{genly.leon@ucn.cl}}
\emailAdd{angel.garcia@ibero.mx}
\emailAdd{guillermo.fernandez@ibero.mx}
\emailAdd{ahalmada@uaq.mx}
\emailAdd{juan.magana@ucentral.cl}
\emailAdd{esteban.gonzalez@uac.cl}

\abstract{Recently, a new field of study called fractional cosmology has emerged. It uses fractional calculus to modify the standard derivative equations and change the Friedmann equations. The evolution of cosmic species densities is also affected by the $\mu$ fractional parameter and the age of the Universe $t_0$. This new approach to cosmology modifies the Friedmann equations and allows for a late cosmic acceleration without the need for a dark energy component. This could be a breakthrough in solving longstanding problems in cosmology. By analyzing observational Hubble data and Type Ia supernovae, we have been able to place strict constraints on the fractional and cosmological parameters. Our results suggest that the Universe may be older than previously estimated. We also explore whether fractional cosmology can help resolve the $H_0$ tension.}

\begin{document}
\maketitle
\section{Introduction}
{\it Fractional calculus} is a subject that has been introduced previously with two works by Niels Henrik Abel in 1823 and 1826. It is a natural extension of the calculus discovered independently in the late 17th century by Isaac Newton and Gottfried Leibniz,  where differentiation and integration are extended to noninteger or complex orders. Denoting by $\mu$ the fractional orders of the operations, the new operations coincide with the results of classical calculus when $\mu\in \mathbb{Z}$, and if it is a positive integer (differentiation) or negative integer (integration). Fractional calculus has drawn increasing attention to studying physical behaviours. Research in fractional differentiation is inherently multi-disciplinary and has its application across various disciplines, for example, fractional quantum mechanics and gravity for fractional spacetime, fractional quantum field theory and cosmology \cite{LimEab+2019+237+256, VargasMoniz:2020hve, Moniz:2020emn, El-Nabulsi:2009bup, El-Nabulsi:2013hsa, Roberts:2009ix, Vacaru:2010wn, Shchigolev:2010vh, Shchigolev:2012rp, Shchigolev:2013jq,   Shchigolev:2015rei,  Calcagni:2016ofu, Calcagni:2020ads, Calcagni:2021ipd, Calcagni:2021aap, Shchigolev:2021lbm, Landim:2021www, Garcia-Aspeitia:2022uxz, Gonzalez:2023who, Micolta-Riascos:2023mqo}. 

In references \cite{Garcia-Aspeitia:2022uxz} and \cite{Gonzalez:2023who}, the fractional calculus free parameters are estimated using cosmological data. The analysis from the type Ia supernovae (SNe Ia) data, the observational Hubble parameter data (OHD), and the joint analysis lead to best-fit values for the free parameters. Moreover, these best-fit values are used to calculate the age of the Universe, a current deceleration parameter, and a current matter density parameter. In \cite{Micolta-Riascos:2023mqo}, dynamical systems were used to analyze fractional cosmology for different matter contents, obtaining a late-time accelerating cosmology. Despite the discrepancy between the age of the Universe predicted by the fractional calculus approach and that of globular clusters, it is essential to highlight that fractional cosmology would contribute to the solutions to other problems associated with the $\Lambda$CDM model. For example, the late-time acceleration without dark energy can alleviate the so-called Cosmological Constant problem, in which the observational value of the $\Lambda$ differs between $60$ and $120$ orders of magnitude compared with the value anticipated by particle physics \cite{Weinberg:1988cp}. Another problem related to the Dark Energy (DE) is the coincidence problem, which stipulates that, currently, Dark Matter (DM)  and DE densities are of the same order of magnitude, with a fine-tuning problem associated with the context of $\Lambda$CDM model \cite{Zlatev:1998tr}. 
Another issue that fractional cosmology can alleviate is the Hubble tension. Measurements of the Hubble parameter at the current time, $H_{0}$, exhibit a discrepancy of $5\sigma$ between the observational value obtained from cepheids and SNe Ia from the Hubble Space Telescope (HST) \cite{Riess:2021jrx}, and the one inferred from Planck CMB \cite{Planck:2018vyg}. The first corresponds to model-independent measurements, while the second depends on the $\Lambda$CDM model. According to \cite{Riess:2019cxk}, observational issues like the $H_{0}$ tension are strong evidence that physics beyond $\Lambda$CDM model is required, being the fractional cosmology one of these options. Therefore, following this line, a possible alternative to solve this tension is considering extensions beyond $\Lambda$CDM (see \cite{Motta:2021hvl, DiValentino:2021izs} for a review). In \cite{Garcia-Aspeitia:2022uxz}, some results related to the $H_0$ tension were discussed in the context of fractional cosmology. In the present paper, we discuss the main results of  \cite{Garcia-Aspeitia:2022uxz, Micolta-Riascos:2023mqo, Gonzalez:2023who}. 

\section{Modified Friedmann equations}
In Fractional cosmology, the Friedmann equation for the Friedmann-Lema\^{i}tre-Robertson-Walker (FLRW) metric in a flat Universe is modified according to  \cite{Shchigolev:2010vh}
\begin{equation}
   3 H^2+ 3(1-\mu) H t^{-1}= \sum_i\rho_i, \label{FriedmannFrac}
\end{equation} where  $\mu$ is the fractional constant parameter, $H=\dot{a}/a$ is the Hubble parameter, the dot means derivative with respect the cosmic time, and we use units where $8 \pi G =1 $.

Assuming separated conservation equations, we have \cite{Garcia-Aspeitia:2022uxz}
\begin{equation}
\dot{\rho}_i+3\left(H+(1-\mu)/(3t)\right)(\rho_i+p_i)=0.\label{CE}
\end{equation}
If $\mu=1$, the standard cosmology without $\Lambda$ is recovered. We are assuming the equation of state $p_i=w_i\rho_i$, with $w_i \neq -1$ constants, setting $a(t_0)=1$, where $t_0$ is the current age of the Universe, and denoting by $\rho_{0i}$ the current value of energy density of the $i$-th species, we obtain $\rho_{i}(t)= \rho_{0i} a(t)^{-3 (1+w_i)} \left(t/t_0\right)^{(\mu-1)(1+w_i)}$. 
For $\mu \neq 1$, we have   
\begin{align}
 \sum_i p_i=6 (\mu -3) H t^{-1}+3H^2- 3(\mu -2) (\mu -1) t^{-2}.\label{BIc}
\end{align}
Eliminating  $\sum_i p_i$ and  $\sum_i \rho_i$, we obtain 
\begin{align}
    & \dot{H}= -2 (\mu -4) H t^{-1}-3 H^2+ (\mu -2) (\mu -1)t^{-2}, \label{Riccati}
\end{align}
Defining the dimensionless time variable $\xi=t/t_0$,  we obtain by integrating \eqref{Riccati}, the following: 
\begin{small}
\begin{align}
& E(\xi)= \frac{H(\xi)}{H_0} =\frac{1}{3 \alpha_0 \xi } \left[\frac{9-2 \mu +r}{2}-\frac{c r}{c +\xi^r}\right], \label{(66)}
\\ 
& z(\xi)= -1+\xi^{\frac{1}{6} (2 \mu +r-9)} \sqrt[3]{\frac{{(c+1)}}{{c+\xi^r}}}, \label{ztau} 
\end{align}
\begin{align}
& p(\xi) = \frac{H_0^2 \left(2 (4 \mu -9) r \left(\xi^{2 r}-c^2\right)+r^2 \left(\xi^r-c \right)^2-7 (4 \mu (2 \mu -9)+45) \left(c +\xi^r\right)^2\right)}{12 \alpha_0^2 \xi^2 \left(c +\xi^r\right)^2}, \label{(68)}\\
& \rho(\xi) = \frac{H_0^2 \left(-2 (5 \mu -12) r \left(\xi^{2 r}-c^2 \right)+r^2 \left(\xi^r-c \right)^2+(2 \mu -9) (8 \mu -15)
 \left(c  +\xi^r\right)^2\right)}{12 \alpha_0^2 \xi^2 \left(c +\xi^r\right)^2}, \label{(69)}
\\
& q(\xi)=-\frac{c^2 (2 \mu +r-9) (2 \mu +r-3) +2 c \left(4 \mu ^2-24 \mu +5 r^2+27\right) \xi^r+(-2 \mu +r+3) (-2 \mu +r+9) \xi^{2 r}}{\left((-2 \mu +r+9) \xi^r-c (2 \mu +r-9)\right)^2}, \label{(67)}
\\
& w_{\text{eff}}(\xi)=\frac{2 (4 \mu -9) r \left(\xi^{2 r}-c^2\right)+r^2 \left(\xi^r-c \right)^2-7 (4 \mu  (2 \mu -9)+45)
   \left(c +\xi^r\right)^2}{\left((-2 \mu +r+9) \xi^r-c (2 \mu +r-9)\right) \left((-8 \mu +r+15) \xi^r-c (8 \mu +r-15)\right)}, \label{(70)}
\\
& \Omega_{\text{m}}(\xi)= \frac{(-8 \mu +r+15) \xi^r-c (8 \mu +r-15)}{(-2 \mu +r+9) \xi^r-c (2 \mu +r-9)} \label{(71)},
\end{align}
 \end{small}
where $\alpha(t)=t H$ is the age parameter, $\alpha_0=H_{0}t_0$, $ c = \frac{-2 \mu +r-6 \alpha_0+9}{2 \mu +r+6 \alpha_0-9}$,  and  $r= \sqrt{8 \mu  (2 \mu -9)+105}$. Taking the limit as $\xi \rightarrow \infty$, we  have 
  \begin{align*}
& \lim_{\xi \rightarrow \infty} z(\xi)=-1,  \quad 
\lim_{\xi \rightarrow \infty} E (\xi)=0, \quad 
\lim_{\xi \rightarrow \infty} p(\xi)=0, \quad 
\lim_{\xi \rightarrow \infty} \rho(\xi)=0, 
\\
& \lim_{\xi \rightarrow \infty} q(\xi)
=\frac{-13-2 (\mu -4) \mu +\sqrt{8 \mu  (2 \mu -9)+105}}{2 (\mu -2) (\mu -1)}, \lim_{\xi \rightarrow \infty} \alpha (\xi) =  \frac{1}{6} \left(9-2 \mu +\sqrt{8 \mu  (2 \mu -9)+105}\right)\geq 0,
\end{align*}
 \begin{equation}\label{Ominfty}
\lim_{t \rightarrow \infty}  w_{\text{eff}} (t) = \frac{-7+\sqrt{8 \mu  (2 \mu -9)+105}}{4 (\mu -1)}, \quad 
  \lim_{t \rightarrow \infty}   \Omega_{\text{m}}  (t) = \frac{5-\sqrt{8 \mu  (2 \mu -9)+105}}{2 (\mu -2)}. 
\end{equation}
The main difficulty of this approach is the need to invert \eqref{ztau} to obtain $\xi$ as a function of $z$ because data is in terms of redshift, which is impossible using analytical tools. After all, the equation is a rational one. By introducing the logarithmic independent variable $s= -\ln(1+z)$, with $s\rightarrow -\infty $ as $z\rightarrow \infty$, $s\rightarrow 0$ as $z\rightarrow 0$, and  $s\rightarrow \infty $ as $z\rightarrow -1$, and  given the initial conditions $\alpha(0)=t_0 H_0, t(0)=t_0$, we obtain the initial value problem 
\begin{align}
   & \alpha '(s)= 9-2 \mu -3 \alpha (s)+\frac{(\mu -2) (\mu -1)}{\alpha (s)}, \label{eq(134)}\\
   &t'(s)={t(s)}/{\alpha(s)}. \label{eq(135)}
\end{align}
The equation \eqref{eq(134)} gives a one-dimensional dynamical system. The equilibrium points are 
$T_1: \alpha = \frac{1}{6} \left(9-2 \mu -\sqrt{8 \mu  (2 \mu -9)+105}\right)$, satisfying $\alpha>0$ for $1<\mu <2$, and $T_2: \alpha= \frac{1}{6} \left(9-2 \mu +\sqrt{8 \mu  (2 \mu -9)+105}\right)$,  that satisfies $\alpha>0$ for $\mu\in\mathbb{R}$. $T_1$ is a source whenever it exists and  $T_2$ is always a sink. That is the asymptotic behaviour for large $t$, which is consistent with \cite{Micolta-Riascos:2023mqo}. We introduce the parameter $\epsilon_0$ such that 
\begin{equation}
 \epsilon_0= \frac{1}{2}\lim_{t\rightarrow \infty} \left(\frac{t_0 H_0 - t H}{t H}\right), \quad \alpha_0=  \frac{1}{6} \left(9 -2 \mu +\sqrt{8 \mu  (2 \mu -9)+105}\right)(1+ 2 \epsilon_0). \label{alpha_0}
\end{equation} 
$\epsilon_0$ is a measure of the limiting value of the relative error in the age parameter $t H$ when it is approximated by $t_0 H_0$. 

\section{Methodology and dataset} \label{sec:constraints}

A Bayesian Markov Chain Monte Carlo (MCMC) analysis is performed to constrain the phase-space parameter ${\bm\Theta}=\{h, \Omega_{0m}, \mu\}$ of the fractional cosmology using observational Hubble data OHD, SNe Ia dataset and joint analysis. Under the \texttt{emcee} Python package environment \citep{Foreman_emcee_2013}, after the auto-correlation time criterion warranty the convergence of the chains, a set of 4000 chains with 250 steps each is performed to establish the parameter bounds. Additionally, the configuration for the priors are Uniform distributions allowing vary the parameters in the range $h\in[0.2, 1]$,  $\Omega_{0m}\in[0,1]$ and $\mu \in[1,3]$.
Hence the figure-of-merit for the joint analysis is built through the Gaussian log-likelihood given as $-2\ln(\mathcal{L}_{\rm data})\varpropto \chi^2_{\rm data}$ and $\chi^2_{\rm Joint} = \chi_{\rm CC}^2 + \chi_{\rm SNe Ia}^2$, 
where each term refers to the $\chi^2$-function for each dataset. Now, each piece of data is described. 

\subsection{Cosmic chronometers}

Up to now, a set of 31 points obtained by differential age tools, namely cosmic chronometers (CC), represents the measurements of the Hubble parameter, which is cosmological independent \citep{Moresco:2016mzx}.
In this sense, this sample is useful to bound alternative models to $\Lambda$CDM. Thus, the figure-of-merit function to minimize is given by
\begin{equation} \label{eq:chiOHD}
    \chi^2_{{\rm CC}}=\sum_{i=1}^{31}\left(\frac{H_{th}(z_i)-H_{obs}(z_i)}{\sigma^i_{obs}}\right)^2,
\end{equation}
where the sum runs over the whole sample, and $H_{th}-H_{obs}$ is the difference between the theoretical and observational Hubble parameter at the redshift $z_i$ and $\sigma_{obs}$ is the uncertainty of $H_{obs}$.

\subsection{Type Ia Supernovae }

Ref. \citep{Scolnic:2018} provides 1048 luminosity modulus measurements, known as Pantheon sample, from Type Ia Supernovae, which cover a region $0.01<z<2.3$. Due to this sample, the measurements are correlated, and it is convenient to build the chi-square function as
\begin{equation}\label{eq:chi2SnIa}
    \chi_{\rm SNe Ia}^{2}=a +\log \left( \frac{e}{2\pi} \right)-\frac{b^{2}}{e},
\end{equation}
where
\begin{equation}
    a = \Delta\boldsymbol{\tilde{\mu}}^{T}\cdot\mathbf{Cov_{P}^{-1}}\cdot\Delta\boldsymbol{\tilde{\mu}}, \quad 
    b = \Delta\boldsymbol{\tilde{\mu}}^{T}\cdot\mathbf{Cov_{P}^{-1}}\cdot\Delta\mathbf{1}, \quad 
    e = \Delta\mathbf{1}^{T}\cdot\mathbf{Cov_{P}^{-1}}\cdot\Delta\mathbf{1}.
\end{equation}
Furthermore, $\Delta\boldsymbol{\tilde{\mu}}$ is the vector of residuals between the theoretical distance modulus and the observed one, $\Delta\mathbf{1}=(1,1,\dots,1)^T$, $\mathbf{Cov_{P}}$ is the covariance matrix formed by adding the systematic and statistic uncertainties, i.e.   $\mathbf{Cov_{P}}=\mathbf{Cov_{P, sys}}+\mathbf{Cov_{P, stat}}$. The super-index $T$ on the above expressions denotes the transpose of the vectors.

The theoretical distance modulus is estimated by
\begin{equation}
    m_{th}=\mathcal{M}+5\log_{10}\left[\frac{d_L(z)}{10\, pc}\right],
\end{equation}
where $\mathcal{M}$ is a nuisance parameter which has been marginalized by Eq. \eqref{eq:chi2SnIa}. 

The luminosity distance, denoted as $d_L(z)$, is computed through  
\begin{equation}\label{eq:dL}
    d_L(z)=(1+z)c\int_0^z\frac{dz^{\prime}}{H(z^{\prime})},
\end{equation}
being $c$ the speed of light.

In  \cite{Garcia-Aspeitia:2022uxz}, the theoretical $H(z)$ is obtained by solving numerically the system 
\begin{small}
\begin{align}
     E(z)=&  -f F(z) 
   + F(z)^{-\mu } \Bigg\{ f^2 F(z)^{2 (\mu +1)}   +\Omega_{0\text{m}} (z+1)^3
   F(z)^{\mu +1}   +\Omega_{0\text{r}} (z+1)^4 F(z)^{\frac{2 (\mu +2)}{3}}\Bigg\}^{1/2}, \label{FriedmannFinal1}
  \\
     F'(z)  = &\frac{2 f F(z)^{\mu +2}}{(\mu -1) (z+1)}  
     \Bigg\{f F(z)^{\mu +1} 
    -\Bigg[  f^2 F(z)^{2 \mu
   +2}  +\Omega_{0\text{m}} (z+1)^3 F(z)^{\mu +1} 
       +\Omega_{0\text{r}} (z+1)^4 F(z)^{\frac{2 (\mu +2)}{3}}\Bigg]^{1/2}\Bigg\}^{-1}, \label{F'(z)}
\end{align}
\end{small}
where $f\equiv(1-\mu)/(2 t_0H_0)$ is going to be the \textit{fractional constant} that will act as the cosmological constant. For $\mu >1$, $\Omega_{0\text{m}}+\Omega_{0\text{r}}<1$, and notice that we choose the positive branch in order to have $E(z)>0$ and where $\Omega_{0\text{r}}=2.469\times10^{-5}h^{-2}(1+0.2271N_{\text{eff}})$, where $N_{\text{eff}}=2.99\pm0.17$ \citep{Planck:2018vyg}.

On the other hand, in reference \cite{Gonzalez:2023who} was constrained the free parameters with the SNe Ia data and OHD using more data points. In particular, the first one has used the same sample as in \cite{Garcia-Aspeitia:2022uxz}. The second one is considered the OHD sample compiled by Magaña \textit{et al.} \cite{Magana_Cardassian_2018}, which consists of $51$ data points from cosmic chronometers and BAO estimations in the redshift range $0.07\leq z\leq 2.36$. 
Hence, for the constraint, we numerically integrate the system given by Eqs. \eqref{eq(134)} and \eqref{eq(135)}, which represent a system for the variables $(\alpha,t)$ as a function of $s=-\ln{\left(1+z\right)}$, and for which we consider the initial conditions $\alpha(s=0)\equiv\alpha_{0}=t_{0}H_{0}$ and $t(s=0)\equiv t_{0}=\alpha_{0}/H_{0}$. Then, the Hubble parameter is obtained numerically by $H_{th}(z)=\alpha(z)/t(z)$. For this integration, we consider the NumbaLSODA code, a python wrapper of the LSODA method in ODEPACK to C++ \footnote{Available online in the GitHub repository \url{https://github.com/Nicholaswogan/numbalsoda}}. For further comparison, were also constrained the free parameters of the $\Lambda$CDM model by excluding radiation, whose respective Hubble parameter as a function of the redshift is given by $H(z)=H_{0}\sqrt{\Omega_{m,0}(1+z)^{3}+1-\Omega_{m,0}}$.

Finally, the free parameter $\alpha_{0}$ was considered through the parameterization given by Eq. \eqref{alpha_0}. Therefore, the free parameters of the Fractional cosmological model are $\theta=\{h,\mu,\epsilon_{0}\}$, and the free parameters of the $\Lambda$CDM model are $\theta=\{h,\Omega_{m,0}\}$. For the free parameters $\mu$, $\epsilon_{0}$, and $\Omega_{m,0}$ were considered the following flat priors: $\mu\in F(1,4)$, $\epsilon_{0}\in F(-0.1,0.1)$, and $\Omega_{m,0}\in F(0,1)$. On the other hand, the prior chosen for $\epsilon_{0}$ is because $\epsilon_0$ is a measure of the limiting value of the relative error in the age parameter $t H$ when it is approximated by $t_0 H_0$ as given by Eq. \eqref{alpha_0}. For the mean value $\epsilon_0=0$, we acquire $\alpha_0=\frac{1}{6} (-2 \mu +r+9)$, and then, we have  the leading term  for $E(z)\simeq (1+z)^{\frac{6}{ (9-2 \mu +r)}}$. The lower prior of $\mu$ is because the Hubble parameter $H \simeq (\mu-1)/t$ becomes negative when $\mu<1$, in the absence of matter.

\subsection{Results and discussion}
\label{Sect:VII.4}

\begin{figure}
    \centering
    \includegraphics[width=0.45\columnwidth]{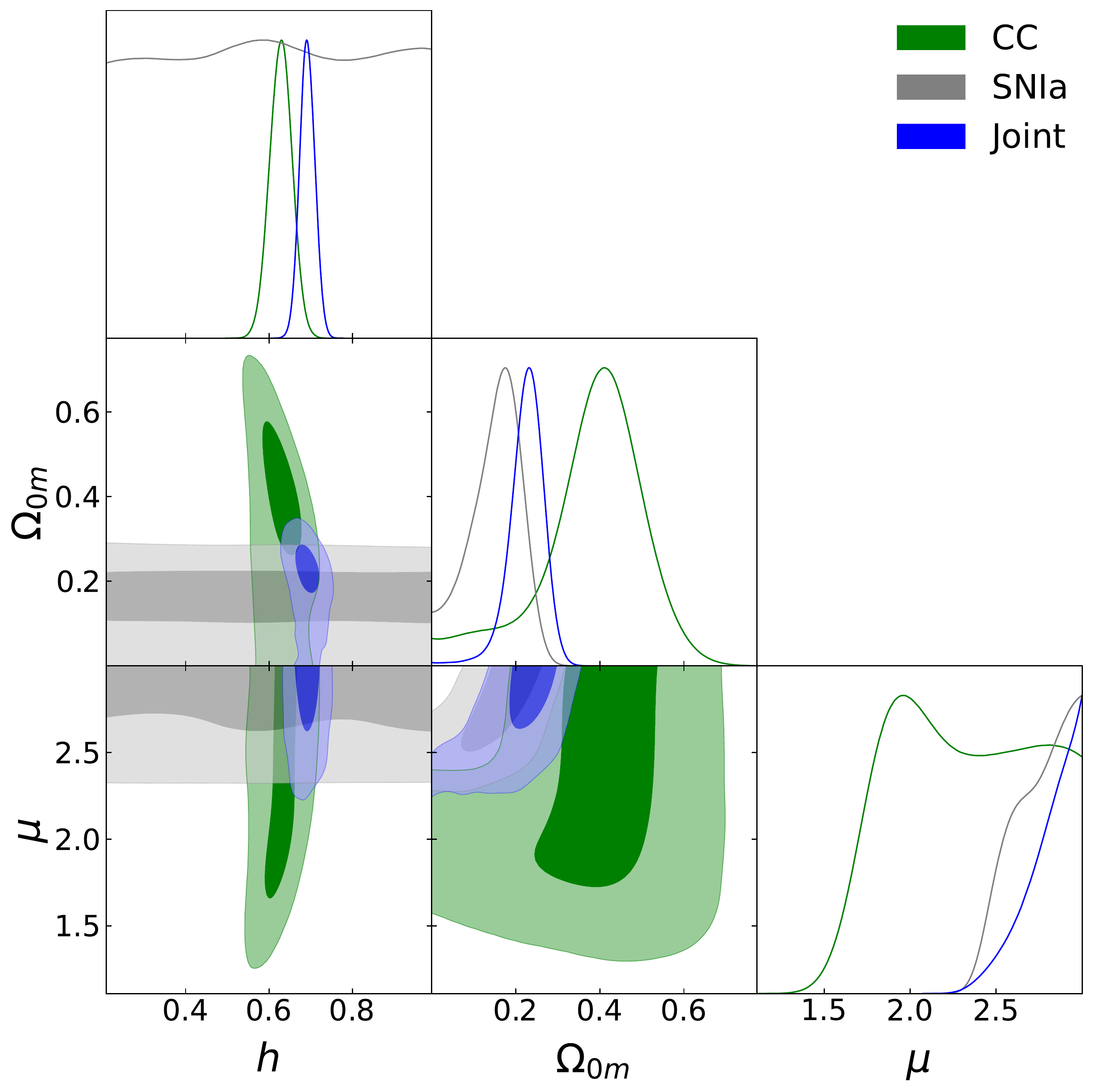}
    \caption{{2D likelihood contours at 68\% and 99.7\% CL, alongside the corresponding 1D posterior distribution of the free parameters (see Ref. \cite{Garcia-Aspeitia:2022uxz})}}
    \label{fig:contours}
\end{figure}

\begin{figure}[h]
     \begin{subfigure}[b]{0.45\textwidth}
         \centering
         \includegraphics[width=\textwidth]{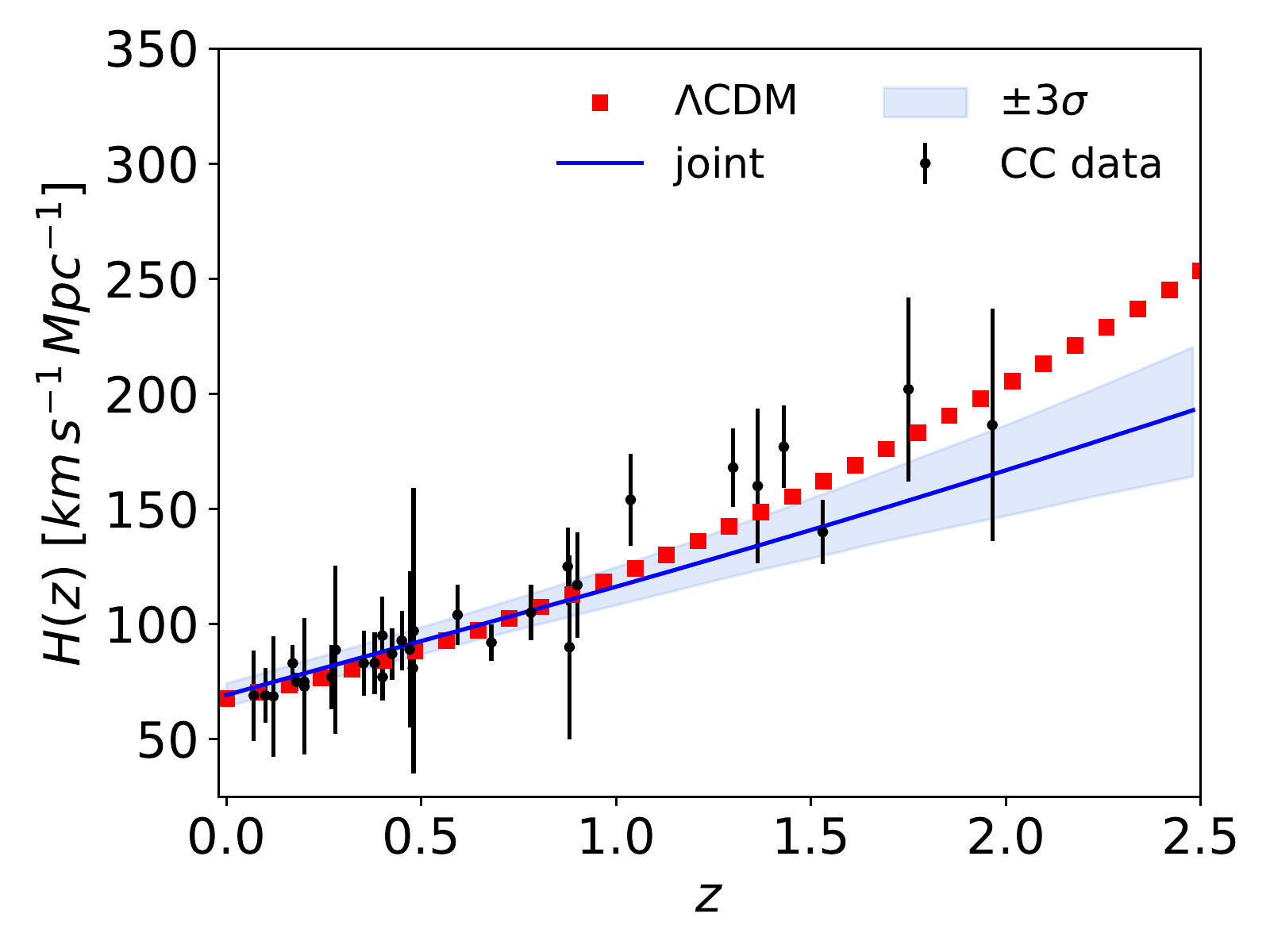}
         \caption{Reconstruction of the $H(z)$ in fractional cosmology}
        \label{fig:Hz}
     \end{subfigure}
     \hfill
     \begin{subfigure}[b]{0.45\textwidth}
         \centering
         \includegraphics[width=\textwidth]{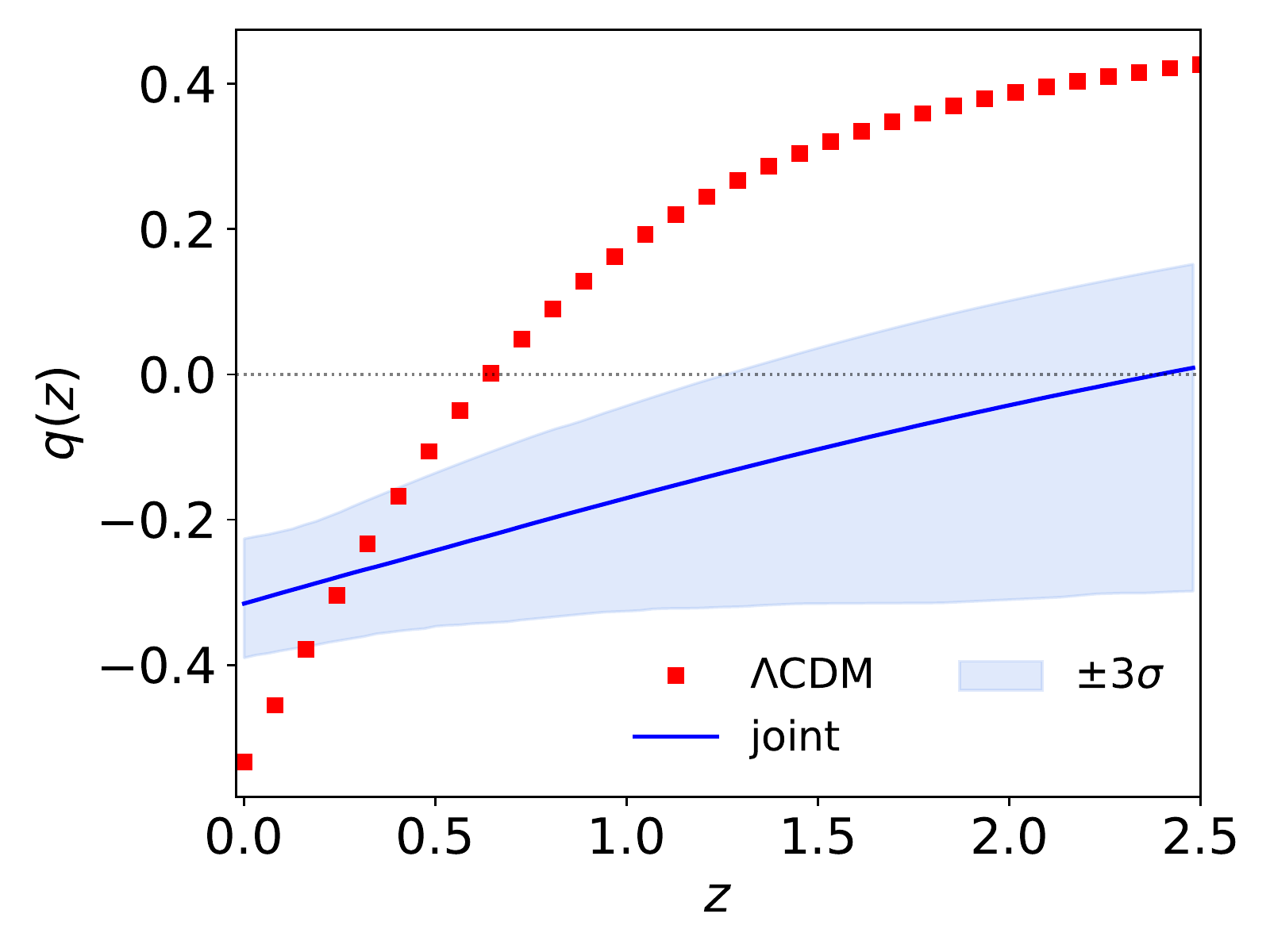}
         \caption{Reconstruction of the $q(z)$ in fractional cosmology}
        \label{fig:qz}
     \end{subfigure}
      \begin{subfigure}[b]{0.45\textwidth}
         \centering
         \includegraphics[width=\textwidth]{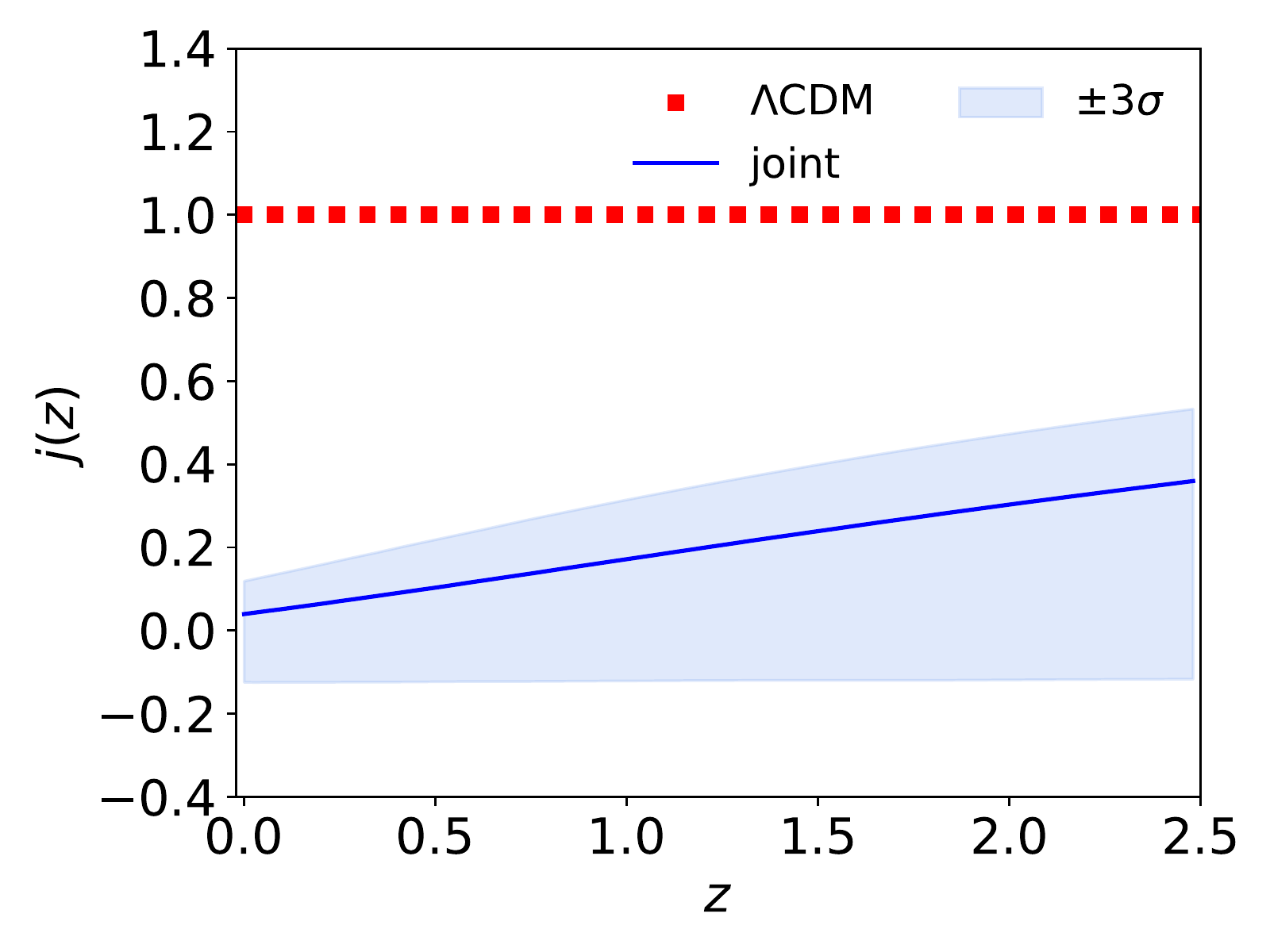}
         \caption{Reconstruction of the $j(z)$ in fractional cosmology}
        \label{fig:jz}
     \end{subfigure}
     \hfill
     \begin{subfigure}[b]{0.45\textwidth}
         \centering
         \includegraphics[width=\textwidth]{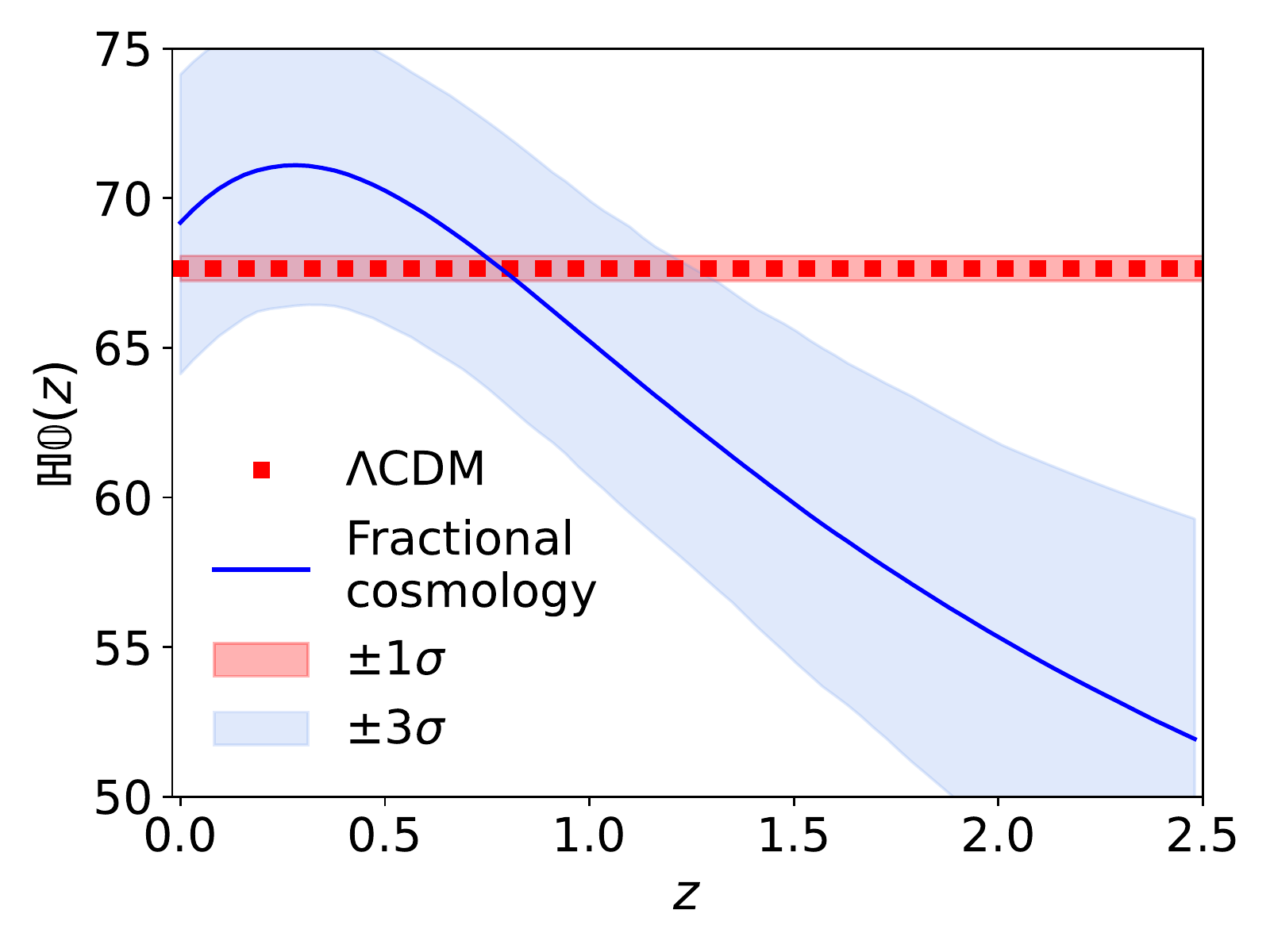}
         \caption{$\mathbf{\mathbb{H}}0(z)$ diagnostic for fractional cosmology}
        \label{JMfig:H0diagnostic}
     \end{subfigure}
     \caption{Reconstruction of Hubble Factor $H(z)$, deceleration parameter $q(z)$, Jerk $j(z)$, and $\mathbf{\mathbb{H}}0(z)$ diagnostic and its comparison against $\Lambda$CDM model (red dashed lines)  with $h=0.6766$ and $\Omega_{\text{m}0}=0.3111$ \cite{Planck:2018vyg} (see Ref. \cite{Garcia-Aspeitia:2022uxz})}
       \end{figure}

The constraints obtained in \cite{Garcia-Aspeitia:2022uxz} through cosmic chronometers, Type Ia Supernovae, and joint analysis are summarized in Fig. \ref{fig:contours} and Table \ref{tab:bestfits} (middle rows). The fractional parameter prefers $\mu=2.839^{+0.117}_{-0.193}$ for a joint analysis which suggests a solid presence of fractional calculus in the dynamical equations of cosmology; however, it generates crucial differences as it is possible to observe from Figs \ref{fig:Hz}, \ref{fig:qz},  and \ref{fig:qz}. From one side, the term $(1-\mu)H/t$ acts like an extra source of mass, closing the Universe and not allowing the observed dynamics, in particular, the Universe acceleration at late times if $\mu<2$, but, for $\mu>2$, we can have an accelerated power-law solution. Furthermore, from Figs. \ref{fig:Hz}, \ref{fig:qz},  and \ref{fig:qz} it is possible to notice that the fractional constant  $f$  can act like the object that causes the Universe acceleration. It is possible to observe from $H(z)$ and $q(z)$ essential differences when we compare them with the standard model, mainly at high redshifts. In addition, the Jerk parameter also shows that the source of the Universe acceleration is not a cosmological constant because, at $z=0$, the fractional parameter does not converge to $j=1$; this is in agreement with recent studies that suggest that $\Lambda$  is not the cause of the Universe acceleration \citep{Zhao:2017cud}.

Moreover, the Universe's age obtained under this scenario is $t_0=33.617^{+3.411}_{-4.511}$ Gyrs based on our Joint analysis, around $2.4$ times larger than the age of the Universe expected under the standard paradigm. However, this value does not contradict the minimum bound expected for the universe age imposed by globular clusters. As far as we know, the maximum bound does not exist and is model-dependent. Fig.  \ref{JMfig:H0diagnostic} displays the reconstruction of the $\mathbf{\mathbb{H}}0(z)$ diagnostic \cite{H0diagnostic:2021} for the fractional cosmology and its error band at $3\sigma$ of confidence level (CL).

\begin{table}[ht!]
    \caption{Best-fit values and $\chi^{2}_{min}$ criteria for the $\Lambda$CDM model with free parameters $h$ and $\Omega_{m,0}$; for the Fractional cosmological model (dust + radiation) \cite{Garcia-Aspeitia:2022uxz}; and the Fractional cosmological model with free parameters $h$, $\mu$, and $\epsilon_{0}$ \cite{Gonzalez:2023who}. The MCMC analysis obtained the values from the SNe Ia data, OHD (or CC), and their joint analysis. The $\Lambda$CDM model is used as a reference model.}
      \setlength{\tabcolsep}{4mm}
      \resizebox{\textwidth}{!}{
    \begin{tabular}{cccccc}
        \hline
         & \multicolumn{4}{c}{\textbf{Best-fit values}} & \\ \\
         \cline{2-5}
         \\
        \textbf{Data} & \boldmath{$h$} & \boldmath{$\Omega_{m,0}$} & \boldmath{$\mu$} & \boldmath{$\epsilon_{0}\times 10^{2}$} & \boldmath{$\chi_{min}^{2}$} \\
        \hline
        \multicolumn{6}{c}{\boldmath{$\Lambda$}\textbf{CDM model}} \\ \\
        \textbf{SNe Ia} & $0.692_{-0.120\;-0.278\;-0.292}^{+0.209\;+0.296\;+0.307}$ & $0.299_{-0.021\;-0.042\;-0.059}^{+0.022\;+0.046\;+0.068}$ & $\cdots$ & $\cdots$ & $1026.9$ \\ \\
        \textbf{OHD} & $0.706_{-0.012\;-0.024\;-0.036}^{+0.012\;+0.024\;+0.035}$ & $0.259_{-0.017\;-0.033\;-0.047}^{+0.018\;+0.038\;+0.059}$ & $\cdots$ & $\cdots$ & $27.5$ \\ \\
        \textbf{SNe Ia+OHD} & $0.696_{-0.010\;-0.020\;-0.029}^{+0.010\;+0.020\;+0.029}$ & $0.276_{-0.014\;-0.027\;-0.040}^{+0.014\;+0.030\;+0.043}$ & $\cdots$ & $\cdots$ & $1056.3$ \\ \\
        \hline
        \multicolumn{6}{c}{\textbf{Fractional cosmological model (dust + radiation) \cite{Garcia-Aspeitia:2022uxz} (The uncertainties presented correspond to $1\sigma(68.3\%)$ CL)}} \\ \\
        \textbf{SNe Ia} & $0.599^{+0.275}_{-0.269}$ & $0.160^{+0.050}_{-0.072}$ & $2.771^{+0.161}_{-0.214}$ & $\cdots$ & $54.83$ \\ \\
        \textbf{CC} & $0.629^{+0.027}_{-0.027}$ & $0.399^{+0.093}_{-0.122}$ & $2.281^{+0.492}_{-0.433}$ & $\cdots$ & $16.14$ \\ \\
        \textbf{SNe Ia+CC} & $0.692^{+0.019}_{-0.018}$ & $0.228^{+0.035}_{-0.040}$ & $2.839^{+0.117}_{-0.193}$ & $\cdots$ & $78.69$ \\ \\
        \hline
        \multicolumn{6}{c}{\textbf{Fractional cosmological model \cite{Gonzalez:2023who} (The uncertainties presented correspond to $1\sigma(68.3\%)$, $2\sigma(95.5\%)$, and $3\sigma(99.7\%)$ CL)}} \\ \\
        \textbf{SNe Ia} & $0.696_{-0.204\;-0.284\;-0.295}^{+0.215\;+0.293\;+0.302}$ & $\cdots$ & $1.340_{-0.245\;-0.328\;-0.339}^{+0.492\;+2.447\;+2.651}$ & $1.976_{-0.905\;-1.848\;-2.067}^{+0.599\;+1.133\;+1.709}$ & $1028.1$ \\ \\
        \textbf{OHD} & $0.675_{-0.008\;-0.015\;-0.021}^{+0.013\;+0.029\;+0.041}$ & $\cdots$ & $2.239_{-0.457\;-0.960\;-1.190}^{+0.449\;+0.908\;+1.386}$ & $0.865_{-0.407\;-0.657\;-0.773}^{+0.395\;+0.650\;+0.793}$ & $29.7$ \\ \\
        \textbf{SNe Ia+OHD} & $0.684_{-0.010\;-0.020\;-0.027}^{+0.011\;+0.021\;+0.031}$ & $\cdots$ & $1.840_{-0.298\;-0.586\;-0.773}^{+0.343\;+1.030\;+1.446}$ & $1.213_{-0.310\;-0.880\;-1.057}^{+0.216\;+0.383\;+0.482}$ & $1061.1$ \\ \\
        \hline
    \end{tabular}}
    \label{tab:bestfits}
\end{table}

\begin{figure}
     \begin{subfigure}[b]{0.45\textwidth}
         \centering
         \includegraphics[width=\textwidth]{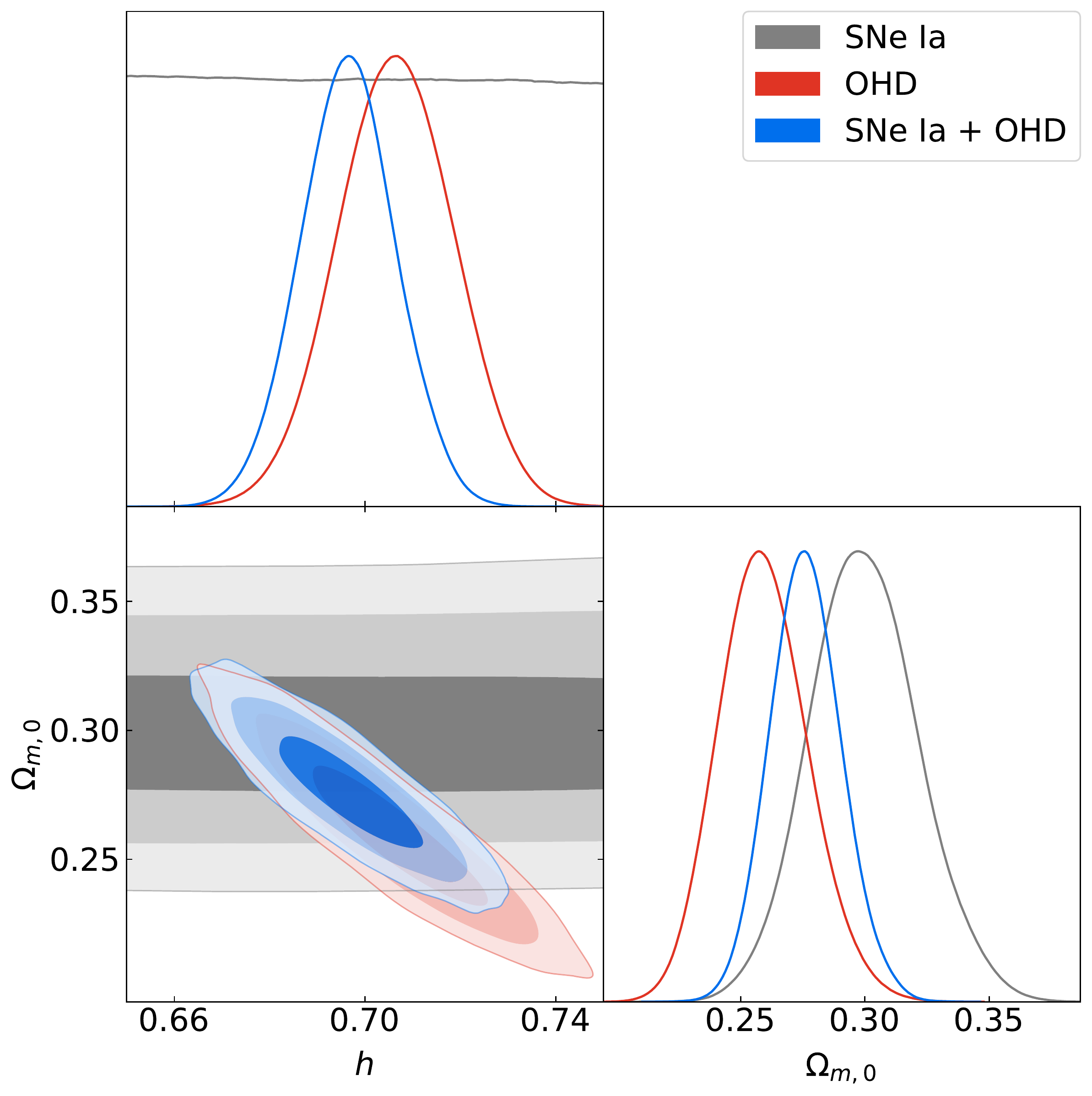}
         \caption{Posterior distribution and joint admissible regions of the free parameters $h$ and $\Omega_{m,0}$ for the $\Lambda$CDM model, obtained in the MCMC analysis.}
        \label{fig:TriangleLCDM}
     \end{subfigure}
     \hfill
     \begin{subfigure}[b]{0.45\textwidth}
         \centering
         \includegraphics[width=\textwidth]{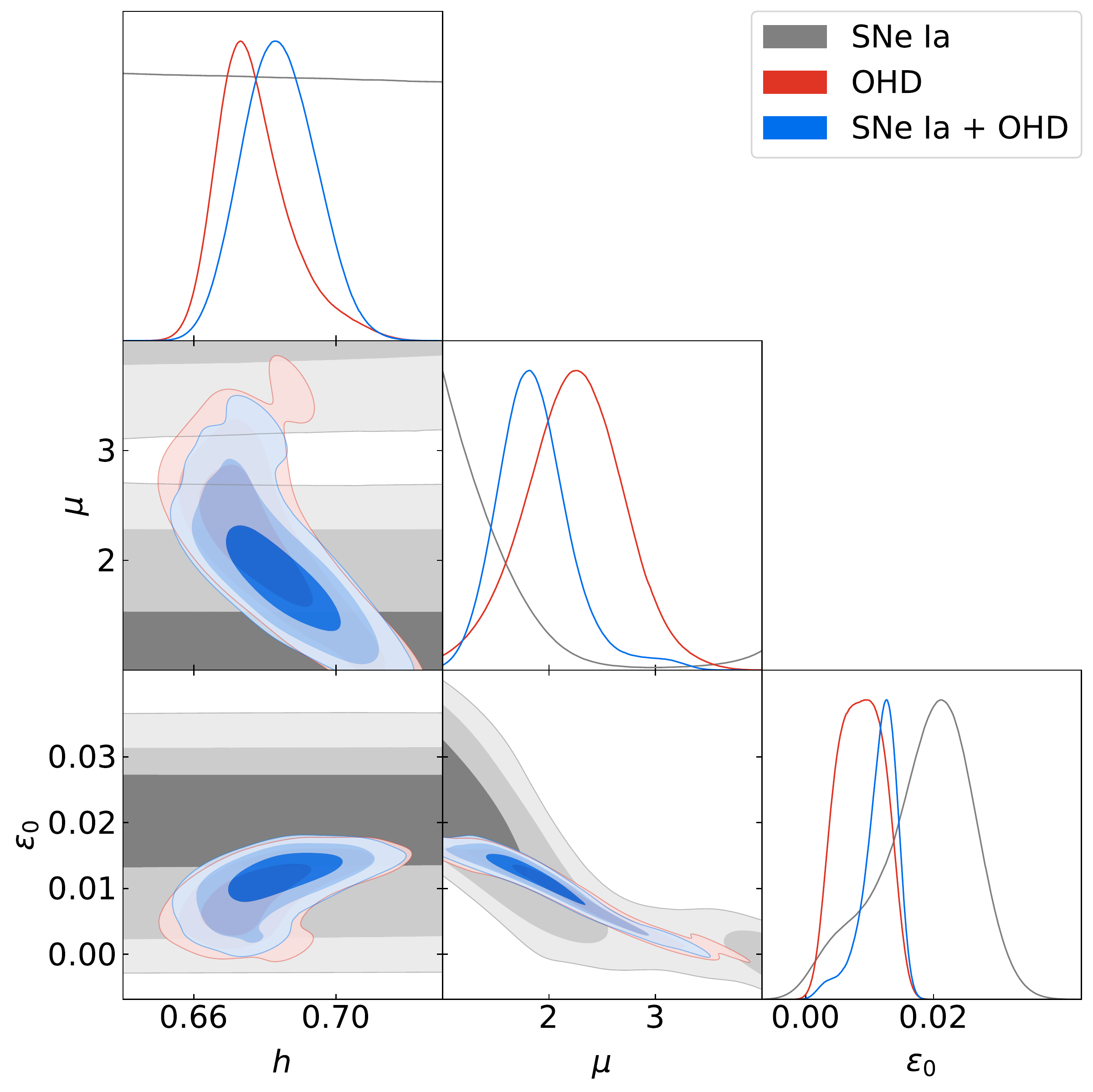}
         \caption{Posterior distribution and joint admissible regions of the free parameters $h$, $\mu$, and $\epsilon_{0}$ for the Fractional cosmological model, obtained in the MCMC analysis.}
        \label{fig:TriangleFractional}
     \end{subfigure}
     \caption{Posterior distribution and joint admissible regions of the free parameters obtained in the MCMC analysis for each model. The admissible joint regions correspond to $1\sigma(68.3\%)$, $2\sigma(95.5\%)$, and $3\sigma(99.7\%)$ CL, respectively. The best-fit values for each model free parameter are shown in Table \ref{tab:bestfits} (see Ref. \cite{Gonzalez:2023who}).}
       \end{figure}

In reference \cite{Gonzalez:2023who}, the best-fit values of the free parameters space for the $\Lambda$CDM model and the Fractional cosmological model, obtained from the SNe Ia data, OHD, and in their joint analysis, with their corresponding $\chi^{2}_{min}$ criteria, are presented in Table \ref{tab:bestfits} (last rows). The uncertainties correspond to $1\sigma$, $2\sigma$, and $3\sigma$ CL. In Figures \ref{fig:TriangleLCDM} and \ref{fig:TriangleFractional}, we depict the posterior distribution and joint admissible regions of the free parameters space of the $\Lambda$CDM model and the Fractional cosmological model, respectively. The joint admissible regions correspond to $1\sigma$, $2\sigma$, and $3\sigma$ CL. Due to the degeneracy between $H_{0}$ and $\mathcal{M}$, the distribution of $h$ for the SNe Ia data was not represented in their full parameter space. The analysis from the SNe Ia data leads to $h=0.696_{-0.295}^{+0.302}$, $\mu=1.340_{-0.339}^{+2.651}$ and  $\epsilon_0=\left(1.976_{-2.067}^{+1.709}\right)\times 10^{-2}$, which are the best-fit values at $3\sigma$ CL. In this case, the value obtained for $h$ cannot be considered as a best fit due to the degeneracy between $H_{0}$ and $\mathcal{M}$. On the other hand, the lower limit of the best fit for $\mu$ is very close to $1$. That is because the posterior distribution for this parameter is close to this value, as seen from Figure \ref{fig:TriangleFractional}. That indicates that a value of the SNe Ia data prefers $\mu<1$, but, as a reminder, this value leads to a negative Hubble parameter in the absence of matter. However, as can be seen from the same Figure \ref{fig:TriangleFractional}, the posterior distribution for these parameters is multi-modal. Therefore, it is possible to obtain a best-fit value that satisfies $\mu>1$. 

It is important to mention that the OHD and the joint analysis do not experience this issue, which allows us to maintain the validity of the prior used for $\mu$. The analysis from OHD leads to  $h=0.675_{-0.021}^{+0.041}$, $\mu=2.239_{-1.190}^{+1.386}$ and $\epsilon_0=\left(0.865_{-0.773}^{+0.793}\right)\times 10^{-2}$, which are the best-fit values at $3\sigma$ CL. In this case, note how the OHD can properly constrain the free parameters $h$, $\mu$ and $\epsilon_{0}$, i.e., we obtain the best fit for the priors considered in our MCMC analysis. Also, note how the posterior distribution of $\mu$ includes the value of $1$, as seen from Figure \ref{fig:TriangleFractional}, but greater than $3\sigma$ CL.

Finally, the joint analysis with data from SNe Ia + OHD leads to $h=0.684_{-0.027}^{+0.031}$, $\mu=1.840_{-0.773}^{+1.446}$ and $\epsilon_0=\left(1.213_{-1.057}^{+0.482}\right)\times 10^{-2}$, which are the best-fit values at $3\sigma$ CL. Focusing our analysis on these results, we can conclude that the region in which $\mu>2$ is not ruled out by observations. On the other hand, these best-fit values lead to an age of the Universe with a value of $t_0=\alpha_0/H_0=25.62_{-4.46}^{+6.89}\;\text{Gyrs}$ at $3\sigma$ CL. This fact to find a universe roughly twice older as one of the $\Lambda$CDM models, which is also in disagreement with the value obtained with globular clusters, with a value of $t_0=13.5^{+0.16}_{-0.14}\pm 0.23$ \citep{Valcin:2021}, is a distinction of the Fractional Cosmology. This result also agrees with the analysis made in \cite{Garcia-Aspeitia:2022uxz}, section 8, where the best-fit $\mu$-value is obtained from the reconstruction of $H(z)$ for different priors of $\mu$. In \cite{Garcia-Aspeitia:2022uxz} was considered a set of 31 points obtained by differential age tools, namely cosmic chronometers (CC), represents the measurements of the Hubble parameter, which is cosmological independent \cite{Moresco:2016mzx} (in \cite{Gonzalez:2023who} was considered the datasets from \cite{Magana_Cardassian_2018}, which consists of $51$ data points in the redshift range $0.07\leq z\leq 2.36$, 20 more points as compared with \cite{Moresco:2016mzx}; nevertheless, the additional points are model-dependent). The 1048 luminosity modulus measurements, known as the Pantheon sample, from Type Ia Supernovae cover a region $0.01<z<2.3$  \cite{Scolnic_Complete_2018}. In \cite{Garcia-Aspeitia:2022uxz}, results depend on the priors used for $\mu$. Say, for the prior
 $0< \mu<1$, $\mu=0.50$ and $t_0=41.30\; \text{Gyrs}$; for $ 1 < \mu < 3$, $\mu=1.71$ and $t_0=27.89\; \text{Gyrs}$; and for $0 < \mu< 3$, $\mu=1.15$ and $t_0=33.66\; \text{Gyrs}$.

\begin{figure}[h]
     \begin{subfigure}[b]{0.45\textwidth}
         \centering
         \includegraphics[width=\textwidth]{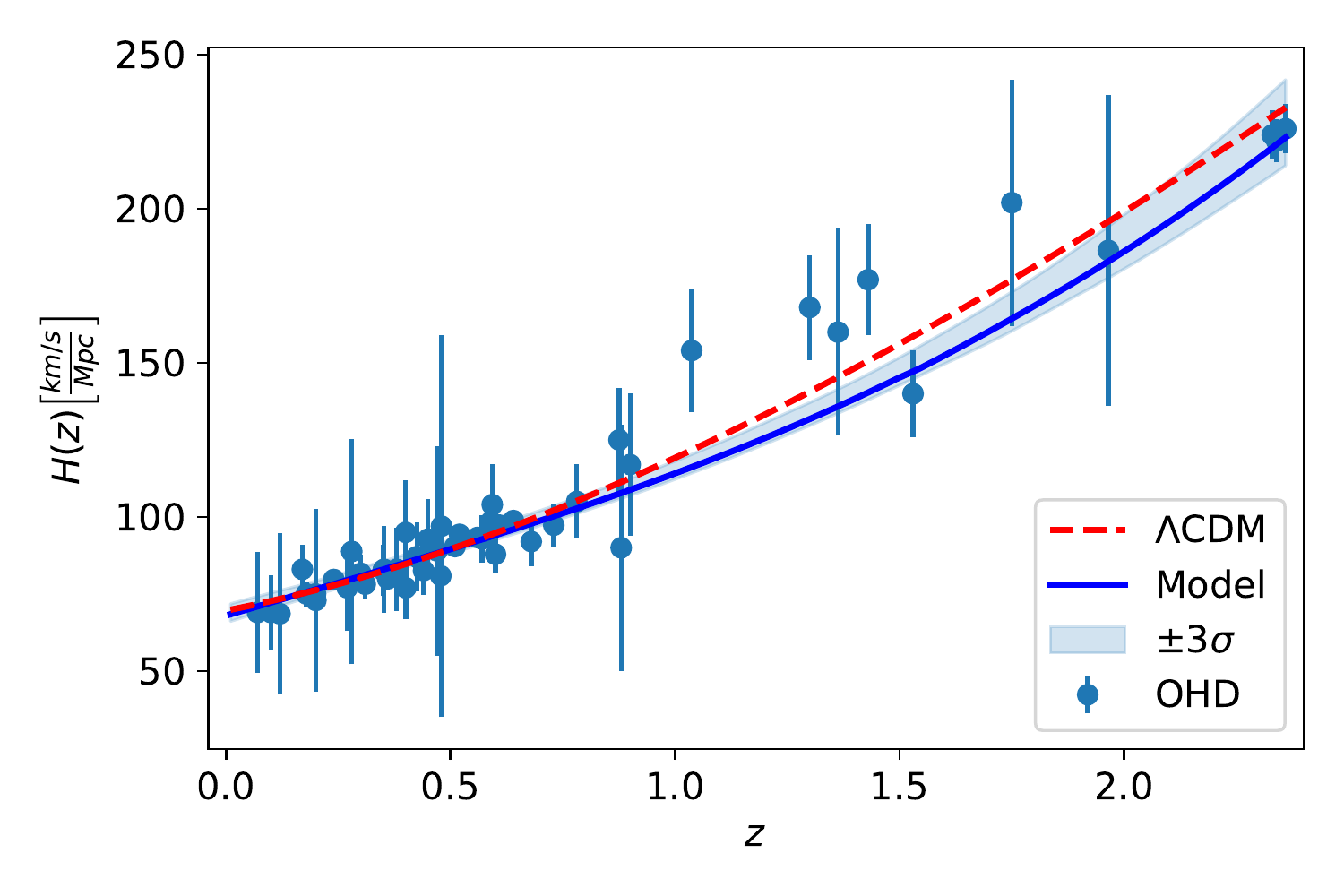}
         \caption{Reconstruction of the $H(z)$}
        \label{fig:OHDFractional}
     \end{subfigure}
     \hfill
     \begin{subfigure}[b]{0.45\textwidth}
         \centering
         \includegraphics[width=\textwidth]{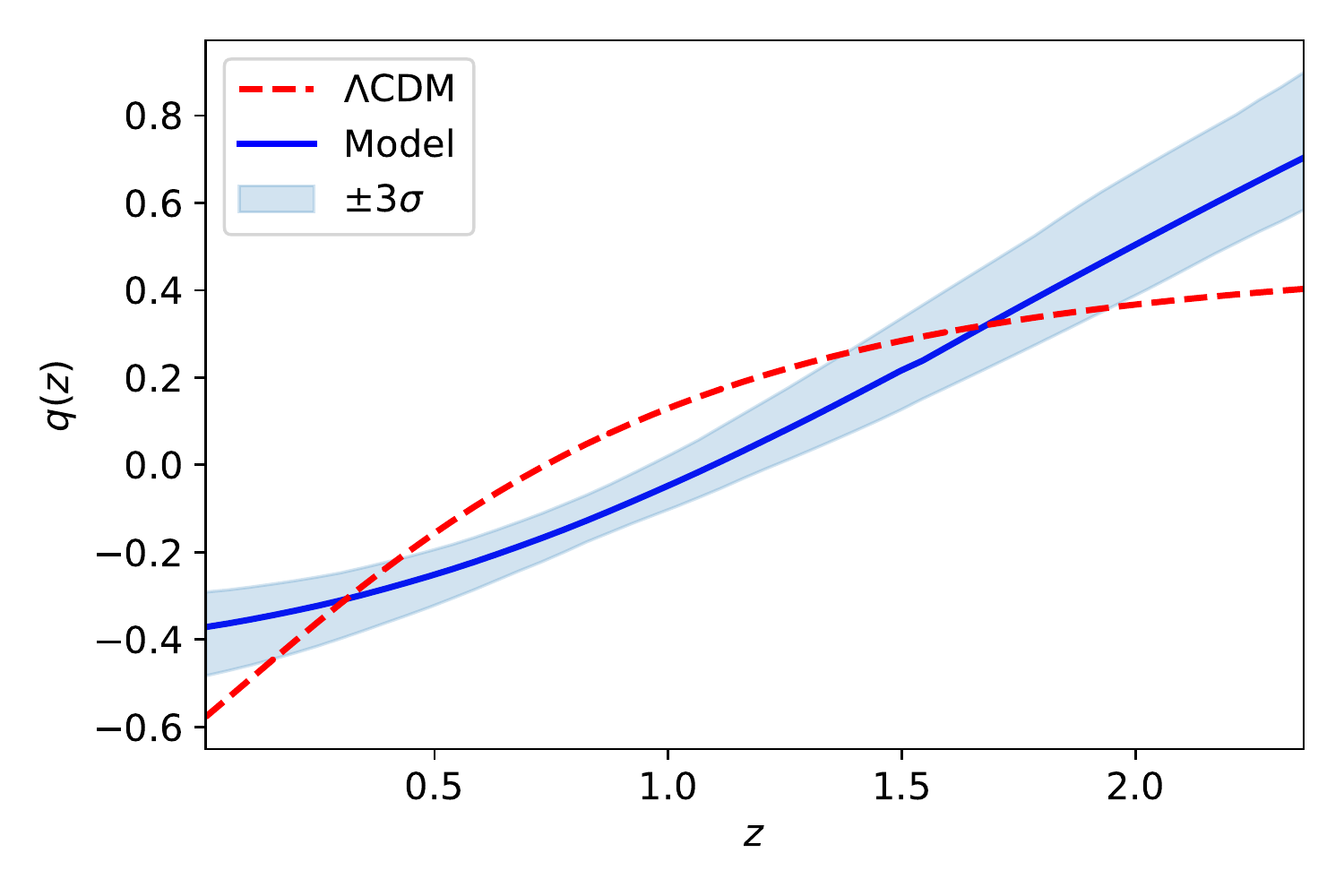}
         \caption{Reconstruction of the $q(z)$}
       \label{fig:qFractional}
     \end{subfigure}
      \begin{subfigure}[b]{0.45\textwidth}
         \centering
         \includegraphics[width=\textwidth]{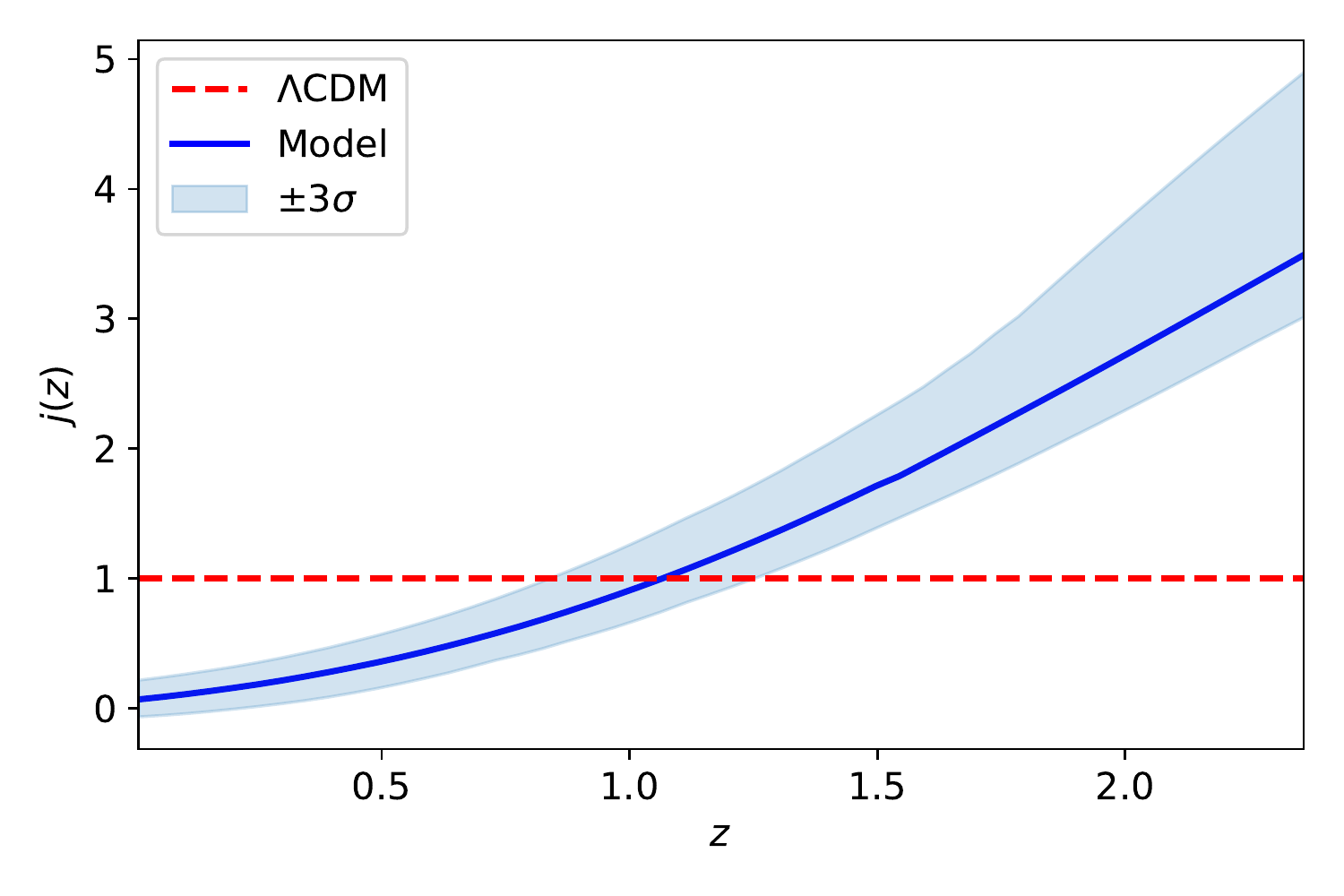}
         \caption{Reconstruction of the $j(z)$}
        \label{fig:JerkFractional}
     \end{subfigure}
     \hfill
     \begin{subfigure}[b]{0.45\textwidth}
         \centering
         \includegraphics[width=\textwidth]{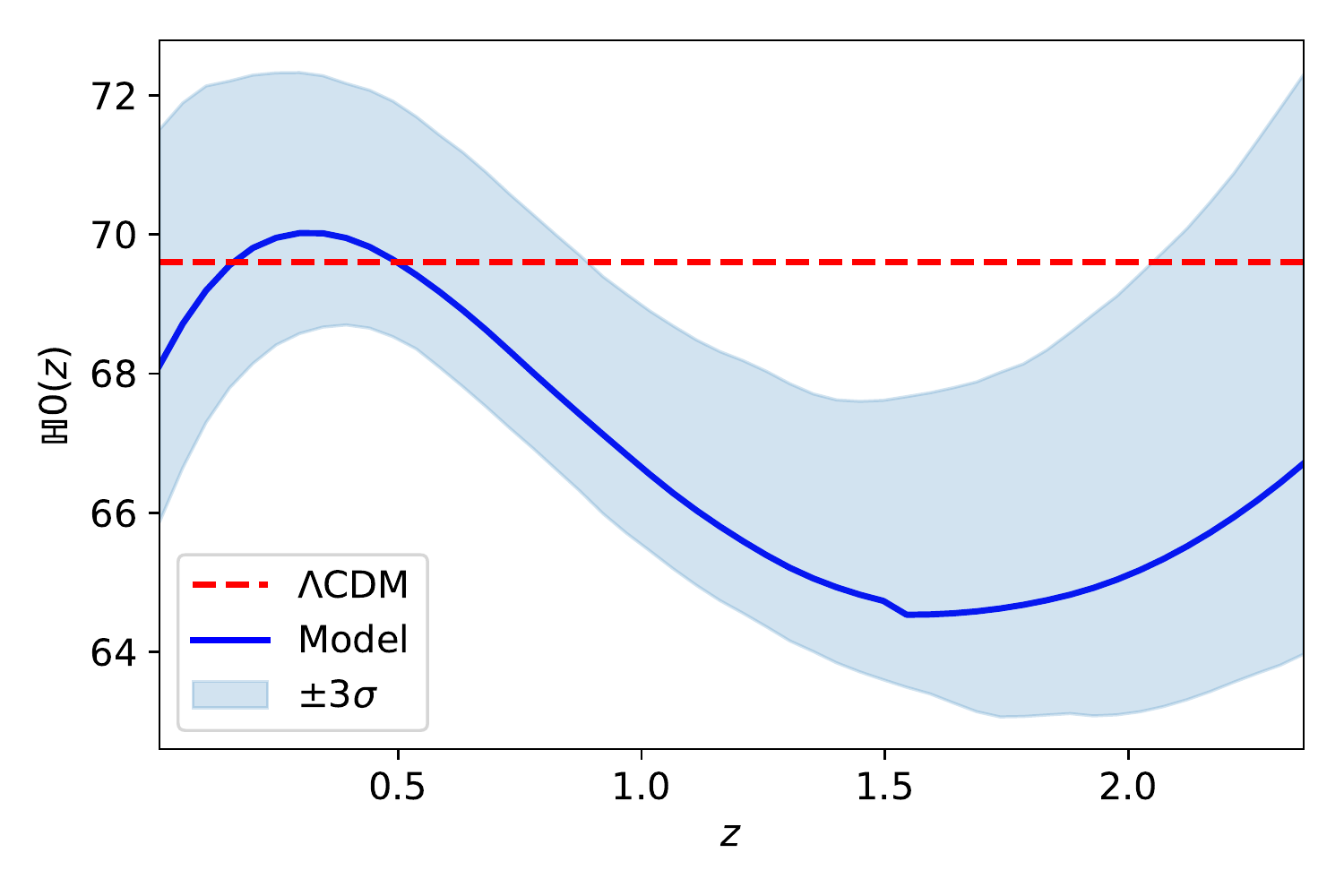}
         \caption{$\mathbf{\mathbb{H}}0(z)$ diagnostic}
       \label{fig:H0diagnostic}
     \end{subfigure}
     \caption{Theoretical Hubble parameter, Deceleration parameter, Jerk,  and $\mathbf{\mathbb{H}}0$ diagnostic (solid blue line)  as a function of the redshift $z$ for the Fractional cosmological model. The shaded curve represents the confidence region at $3\sigma(99.7\%)$ CL. Each model is compared with the $\Lambda$CDM model (red dashed line). Fig. \ref{fig:OHDFractional} is contrasted with the OHD sample. The rest of the figure is obtained using the best-fit values from the joint analysis in Table \ref{tab:bestfits} (see Ref. \cite{Gonzalez:2023who}) }
       \end{figure}

\begin{figure}
    \centering
    \includegraphics[scale = 0.45]{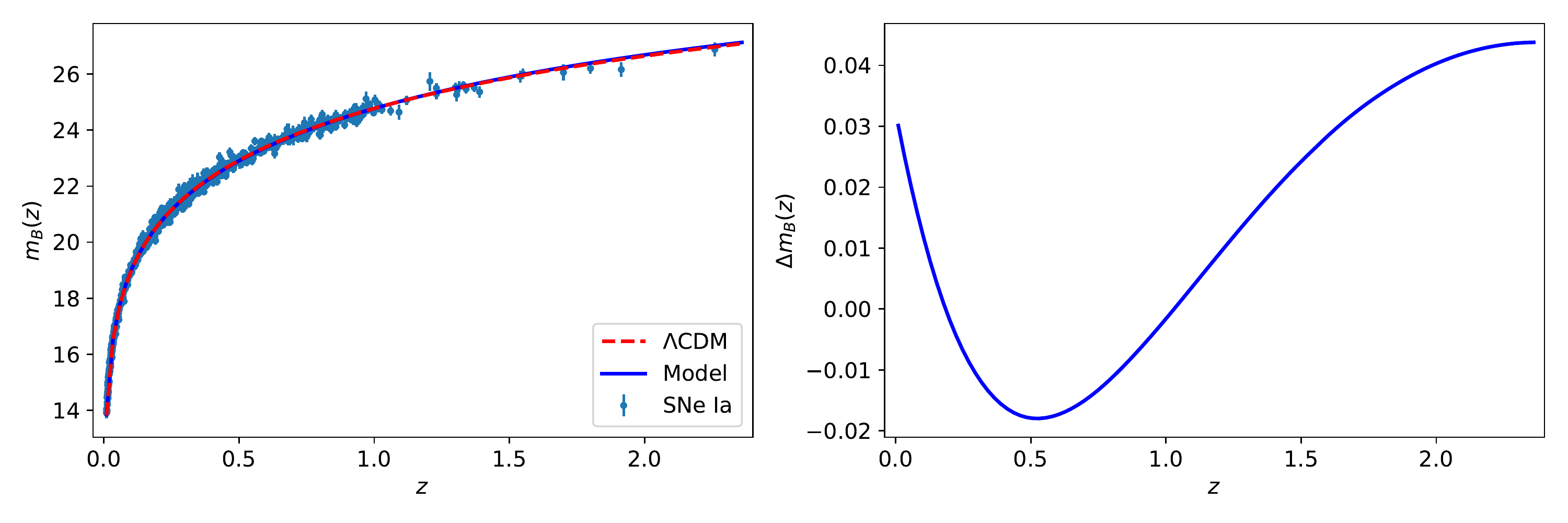}
    \caption{(\textit{left panel}) Theoretical apparent B-band magnitude for the $\Lambda$CDM model (red dashed line) and the Fractional cosmological model (solid blue line) as a function of the redshift $z$, contrasted with the Pantheon data set. (\textit{right panel}) Variation of the theoretical apparent B-band magnitude of the Fractional cosmological model concerning the $\Lambda$CDM model as a function of the redshift $z$. The curve is obtained through the expression $\Delta m_{B}=m_{B, Model}-m_{B,\Lambda CDM}$. The figures are obtained using the best-fit values from the joint analysis in Table \ref{tab:bestfits} (see Ref. \cite{Gonzalez:2023who})}
    \label{fig:SNeFractional}
\end{figure}

From the values for the $\chi^{2}_{min}$ criteria presented in Table \ref{tab:bestfits} (first rows), it is possible to see that the $\Lambda$CDM model is the best model to fit the SNe Ia, OHD, and joint data. Nevertheless, the Fractional cosmological model studied in \cite{Gonzalez:2023who} exhibits values of the $\chi^{2}_{min}$ criteria close to the values of the $\Lambda$CDM model, with differences of $1.2$ for the SNe Ia data, $2.2$ for the OHD data, and $4.8$ for their joint analysis. So, this Fractional cosmological model is suitable for describing the SNe Ia and OHD data, as seen from Figure \ref{fig:OHDFractional} and Figure \ref{fig:SNeFractional}, showing the transition from a deceleration expansion phase to an accelerated one. Therefore, Fractional Cosmology can be considered an alternative valid cosmological model to describe the late-time Universe.

On the other hand, Fig. \ref{fig:OHDFractional} shows the theoretical Hubble parameter for the $\Lambda$CDM model (red dashed line) and the Fractional cosmological model (solid blue line) as a function of the redshift $z$, contrasted with the OHD sample. The shaded curve represents the confidence region of the Hubble parameter for the Fractional cosmological model at $3\sigma(99.7\%)$ CL. The figure is obtained using the best-fit values from the joint analysis in Table \ref{tab:bestfits} (last rows). Additionally, in order to establish that this Fractional cosmological model can describe a universe that experiences a transition from a decelerated expansion phase to an accelerated one, we compute the deceleration parameter $q=-1-\dot{H}/H^{2}$, which using the Riccati Equation \eqref{Riccati}, leads to
\begin{equation}
q(\alpha(s))= 2 + \frac{2 (\mu -4) }{\alpha(s)}-\frac{(\mu -2) (\mu -1)}{\alpha^2(s)}.\label{DecelerationFinal}
\end{equation}
Following this line, in Figure \ref{fig:qFractional}, we depict the deceleration parameter for the Fractional cosmological model as a function of the redshift $z$, obtained for the best-fit values from the joint analysis in the Table \ref{tab:bestfits} (last rows), with an error band at $3\sigma$ CL. We also depict the deceleration parameter for the $\Lambda$CDM model as a reference model. From this figure, we can conclude that the Fractional cosmological model effectively experiences this transition at $z_{t}\gtrapprox 1$, with the characteristic that $z_{t}>z_{t,\Lambda CDM}$, being $z_{t,\Lambda CDM}$ the transition redshift of the $\Lambda$CDM model. Even more, the current deceleration parameter of the Fractional cosmological model is $q_{0}=-0.37_{-0.11}^{+0.08}$ at $3\sigma$ CL. Moreover, we compute the cosmographic parameter known as the Jerk, which quantifies if the Fractional cosmological model tends to $\Lambda$ or its another kind of DE, which can be written as 
\begin{align}
    &j(s)=q(s)(2q(s)+1)-\frac{dq(s)}{ds},
\end{align}
where $q$ is given by Eq. \eqref{DecelerationFinal}. Hence, 
\begin{align}
    &j(\alpha(s))= \frac{12 (\mu -4)}{\alpha (s)}+\frac{(\mu -21) \mu +50}{\alpha
   (s)^2}-\frac{2 (\mu -3) (\mu -2) (\mu -1)}{\alpha (s)^3}+10. \label{Jerk}
\end{align}
Figure \ref{fig:JerkFractional} shows the Jerk for the $\Lambda$CDM model (red dashed line) and the Fractional cosmological model (solid blue line) as a function of the redshift $z$. The figure is obtained using the best-fit values from the joint analysis in Table \ref{tab:bestfits} (last rows) with an error band at $3\sigma$ CL, represented by a shaded region. A departure of more than $3\sigma$ of CL for the current value for $\Lambda$CDM shows an alternative cosmology with an effective dynamical equation of state for the Universe for late times in contrast to $\Lambda$CDM.

On the other hand,  in Figure \ref{fig:H0diagnostic}, we depict $\mathbf{\mathbb{H}}0$ diagnostic for the $\Lambda$CDM model (red dashed line) and the Fractional cosmological model (solid blue line) as a function of the redshift $z$. The figure is obtained using the best-fit values from the joint analysis in Table \ref{tab:bestfits} (last rows), with an error band at $3\sigma$ CL, represented by a shaded region. As a reminder, in both Figures \ref{fig:JerkFractional} and \ref{fig:H0diagnostic}, we also depict the Jerk and the $\mathbf{\mathbb{H}}0$ diagnostic for the $\Lambda$CDM model as a reference model.

\begin{figure}
     \begin{subfigure}[b]{0.45\textwidth}
         \centering
         \includegraphics[width=\textwidth]{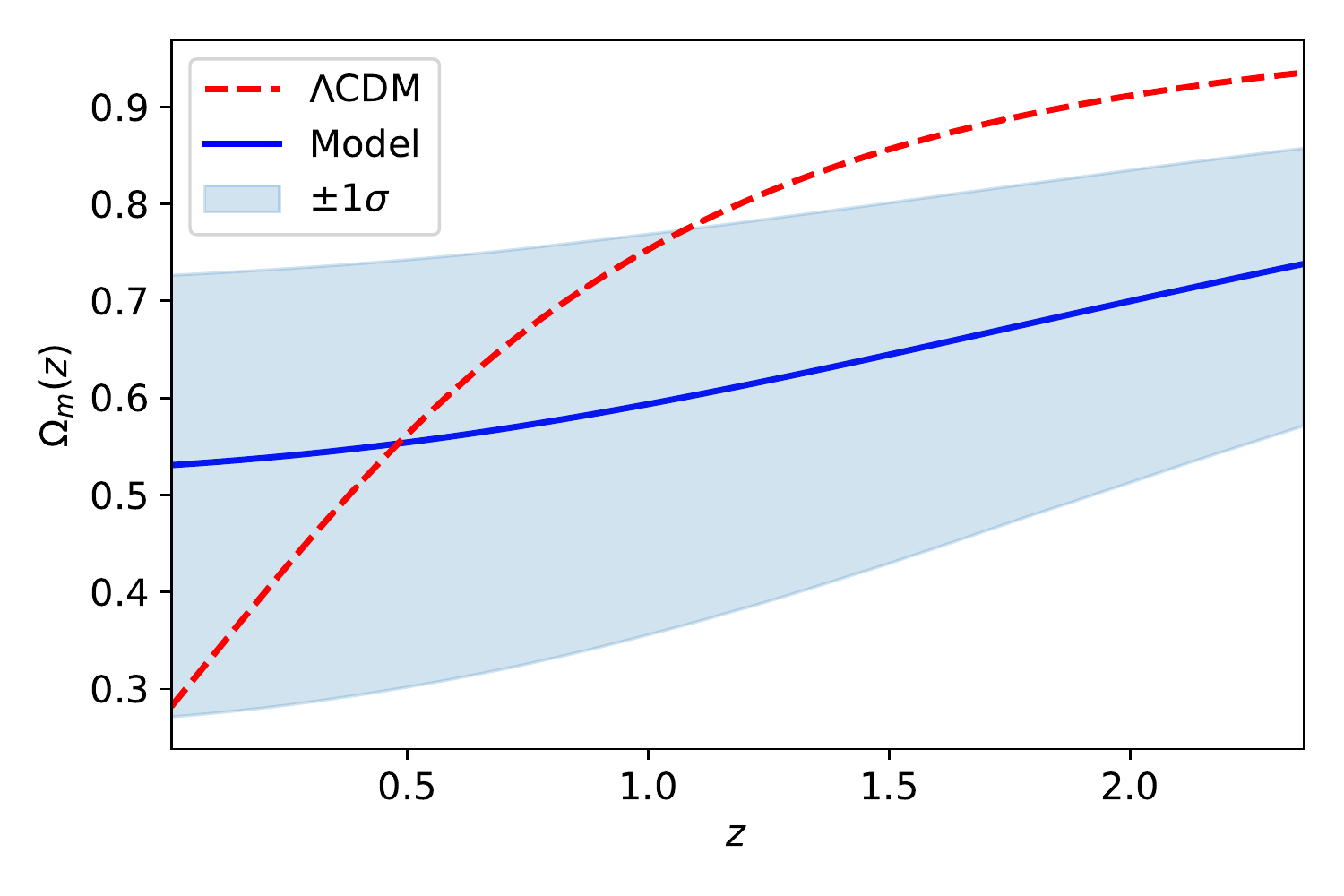}
         \caption{Matter density parameter for the $\Lambda$CDM model (red dashed line) and the Fractional cosmological model (solid blue line) as a function of the redshift $z$. }
        \label{fig:OmFractional}
     \end{subfigure}
     \hfill
     \begin{subfigure}[b]{0.45\textwidth}
         \centering
         \includegraphics[width=\textwidth]{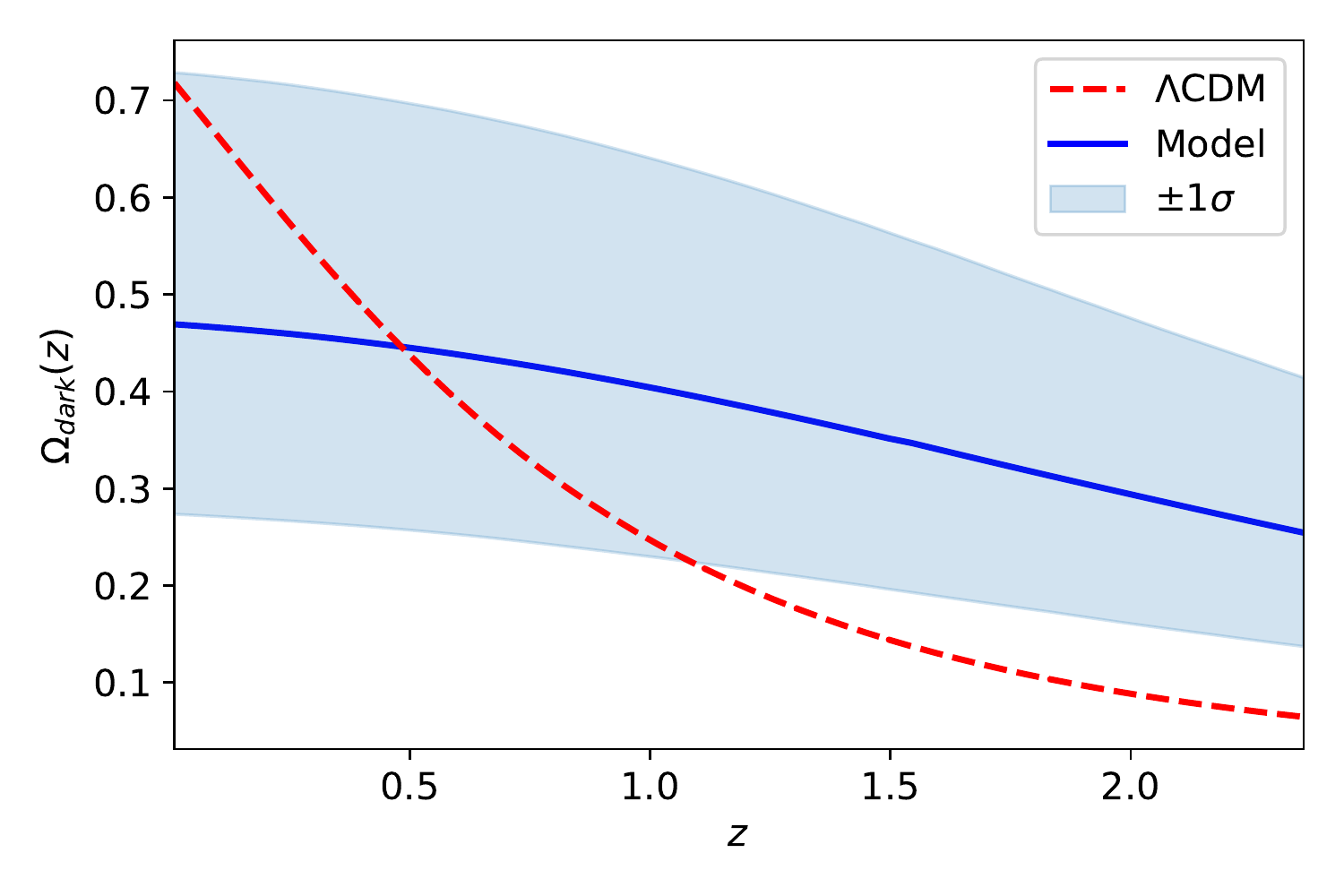}
         \caption{Dark energy density parameter for the $\Lambda$CDM model (red dashed line) and the Fractional cosmological model (solid blue line) as a function of the redshift $z$. }
        \label{fig:OdFractional}
     \end{subfigure}
     \caption{Dark energy and Dark Matter density parameters for the $\Lambda$CDM model and the Fractional cosmological model as a function of the redshift $z$. The shaded curve represents the confidence region of the matter density parameter for the Fractional cosmological model at $1\sigma(68.3\%)$  CL. The figure is obtained using the best-fit values from the joint analysis in Table \ref{tab:bestfits} (see Ref. \cite{Gonzalez:2023who})}
       \end{figure}

In Figures \ref{fig:OmFractional} and \ref{fig:OdFractional}, we depict the matter density and fractional density parameters for the Fractional cosmological model (the last one interpreted as dark energy), respectively, as a function of the redshift $z$, for the best-fit values from the joint analysis in the Table \ref{tab:bestfits}, with an error band at $1\sigma$ CL. We depict the matter density and dark energy density parameters in both figures for the $\Lambda$CDM model. From Figure \ref{fig:OmFractional}, we can see that the matter density parameter for the Fractional cosmological model, obtained from Eq. \eqref{(71)}, presents significant uncertainties, which can be a consequence of their reconstruction from a Hubble parameter that does not take into account any EoS. In this sense, the current value of this matter density parameter at $1\sigma$ CL is $\Omega_{m,0}=0.531_{-0.260}^{+0.195}$, a value that is in agreement with the asymptotic value obtained from Eq. \eqref{Ominfty} of $\Omega_{m,t\to\infty}=0.519_{-0.262}^{+0.199}$, computed at $1\sigma$ CL for the best-fit values from the joint analysis in the Table \ref{tab:bestfits} (last rows). Therefore, this greater value of $\Omega_{m,0}$ for the Fractional cosmological model can, in principle, explain the lower value of the current deceleration parameter $q_{0}$ and the excess of matter in the effective term $\rho_{\text{frac}}=3(\mu-1)t^{-1}H$ with $\Omega_{\text{frac}}(\alpha(s))=(\mu-1)/\alpha(s)$. Note that the current value $\Omega_{\text{frac},0}$ can be interpreted as the dark energy density parameter for the Fractional cosmological model as $\Omega_{\text{frac},0}=0.469_{-0.195}^{+0.260}$, which satisfies the condition $\Omega_{m,0}+\Omega_{\text{frac},0}=1$. Observe that the energy densities of DE and DM are of the same order of magnitude today, alleviating the Coincidence Problem.

Finally, we estimate the free parameters $(\alpha_0, \mu)$ using cosmological data. Using the re-parameterization $H_{0}=100\frac{\text{km/s}}{\text{Mpc}}h$,  $\alpha_0=  \frac{1}{6} \left(9 -2 \mu +\sqrt{8 \mu  (2 \mu -9)+105}\right)(1+ 2 \epsilon_0)$. 

The analysis from the SNe Ia data, OHD and the joint analysis with data from SNe Ia + OHD leads respectively to $h=0.696_{-0.295}^{+0.302}$, $\mu=1.340_{-0.339}^{+2.651}$ and  $\epsilon_0=\left(1.976_{-2.067}^{+1.709}\right)\times 10^{-2}$,   $h=0.675_{-0.021}^{+0.041}$, $\mu=2.239_{-1.190}^{+1.386}$ and $\epsilon_0=\left(0.865_{-0.773}^{+0.793}\right)\times 10^{-2}$, and  $h=0.684_{-0.027}^{+0.031}$, $\mu=1.840_{-0.773}^{+1.446}$ and $\epsilon_0=\left(1.213_{-1.057}^{+0.482}\right)\times 10^{-2}$, where the best-fit values are calculated at $3\sigma$ CL. On the other hand, these best-fit values lead to an age of the Universe with a value of $t_0=\alpha_0/H_0=25.62_{-4.46}^{+6.89}\;\text{Gyrs}$, a current deceleration parameter of $q_{0}=-0.37_{-0.11}^{+0.08}$, both at $3\sigma$ CL, and a current matter density parameter of $\Omega_{m,0}=0.531_{-0.260}^{+0.195}$ at $1\sigma$ CL. Finding a Universe roughly twice older as the one of $\Lambda$CDM is a distinction of Fractional Cosmology. Focusing our analysis on these results, we can conclude that the region in which $\mu>2$ is not ruled out by observations. This parameter region is relevant because, in the absence of matter, fractional cosmology gives a power-law solution $a(t)= \left(t/t_0\right)^{\mu-1}$, which is accelerated for $\mu>2$. In summary, we presented a fractional origin model that leads to an accelerated state without appealing to $\Lambda$ or Dark Energy.

\section{Conclusions}

This paper discusses the formalism of fractional calculus, which
modifies the integer order derivative by a fractional derivative of order $\mu$. It generates changes
in the Friedmann equations, where the standard evolution of the cosmic species densities depends on the fractional
parameter and the Universe's current age $t_0$. The additional term in the new cosmic dynamics equation can support
the late-time accelerated expansion without a dark energy component. We estimated stringent constraints on the fractional
and cosmological parameters using observational Hubble data, Type Ia supernovae and joint analysis to elucidate that. According to our
results, the Universe would be older than the standard estimations. We have obtained modified Friedmann equations at the
background level under fractional calculus, which provides a late cosmic acceleration without
introducing a dark energy component. This radical approach could be a new path to tackle
problems not resolved until now in cosmology. Finally, we analyzed whether fractional cosmology
can alleviate $H_0$ tension. We observe a trend of $H_{0}$ to the value obtained by the Supernova $H_0$ for the Equation of State \citep{Riess:2019cxk} at current times, and in agreement with Planck's value \citep{Planck:2018vyg} for $z\lesssim1.5$. However, a discrepancy between both values in the region $1.5<z<2.5$ holds, so $H_0$ tension is not fully resolved.

\section{Acknowledgments}
GL was funded by  Vicerrectoría de Investigación y Desarrollo Tecnológico (Vridt) at UCN and through Concurso De Pasantías De Investigación Año 2022, Resolución Vridt N 040/2022 under the Project ``The Hubble constant tension: some ways to
alleviate it'' and through Resolución Vridt No. 054/2022. 
M.A.G.-A. acknowledges support from c\'atedra Marcos Moshinsky and Universidad Iberoamericana for support with the SNI grant. GFA acknowledges support from DINVP and Universidad Iberoamericana. AHA thank the support from Luis Aguilar, Alejandro de Le\'on, Carlos Flores, and Jair Garc\'ia of the Laboratorio 
Nacional de Visualizaci\'on Cient\'ifica Avanzada.  J.M. acknowledges the support from  ANID REDES 190147. E. G. acknowledges the support of Direcci\'on de Investigaci\'on y Postgrado at Universidad de Aconcagua.
The authors are thankful for the support of Núcleo de Investigación Geometría Diferencial y Aplicaciones, Resolución VRIDT No. 096/2022. 

\providecommand{\href}[2]{#2}\begingroup\raggedright\endgroup


\begin{thebibliography}{10}

\bibitem{LimEab+2019+237+256}
S.C.~Lim and C.H.~Eab, \emph{Fractional quantum fields},  in \emph{Volume 5
  Applications in Physics, Part B}, V.E.~Tarasov, ed., pp.~237--256, De Gruyter
  (2019), \href{https://doi.org/10.1515/9783110571721-010}{DOI}.

\bibitem{VargasMoniz:2020hve}
P.~V.~Moniz and S.~Jalalzadeh, \emph{{Challenging Routes in Quantum
  Cosmology}}, World Scientific Publishing, Singapore (8, 2020),
  \href{https://doi.org/10.1142/8540}{10.1142/8540}.

\bibitem{Moniz:2020emn}
P.V.~Moniz and S.~Jalalzadeh, \emph{{From Fractional Quantum Mechanics to
  Quantum Cosmology: An Overture}},
  \href{https://doi.org/10.3390/math8030313}{\emph{Mathematics} {\bfseries 8}
  (2020) 313} [\href{https://arxiv.org/abs/2003.01070}{{\ttfamily
  2003.01070}}].

\bibitem{El-Nabulsi:2009bup}
A.R.~El-Nabulsi, \emph{{Fractional Lagrangian formulation of general relativity
  and emergence of complex, spinorial and noncommutative gravity}},
  \href{https://doi.org/10.1142/S021988780900345X}{\emph{Int. J. Geom. Meth.
  Mod. Phys.} {\bfseries 6} (2009) 25}.

\bibitem{El-Nabulsi:2013hsa}
A.R.~El-Nabulsi, \emph{{Fractional derivatives generalization of Einstein`s
  field equations}},
  \href{https://doi.org/10.1007/s12648-012-0201-4}{\emph{Indian J. Phys.}
  {\bfseries 87} (2013) 195}.

\bibitem{Roberts:2009ix}
M.D.~Roberts, \emph{{Fractional Derivative Cosmology}}, {\emph{SOP Trans.
  Theor. Phys.} {\bfseries 1} (2014) 310}
  [\href{https://arxiv.org/abs/0909.1171}{{\ttfamily 0909.1171}}].

\bibitem{Vacaru:2010wn}
S.I.~Vacaru, \emph{{Fractional Dynamics from Einstein Gravity, General
  Solutions, and Black Holes}},
  \href{https://doi.org/10.1007/s10773-011-1010-9}{\emph{Int. J. Theor. Phys.}
  {\bfseries 51} (2012) 1338}
  [\href{https://arxiv.org/abs/1004.0628}{{\ttfamily 1004.0628}}].

\bibitem{Shchigolev:2010vh}
V.K.~Shchigolev, \emph{{Cosmological Models with Fractional Derivatives and
  Fractional Action Functional}},
  \href{https://doi.org/10.1088/0253-6102/56/2/34}{\emph{Commun. Theor. Phys.}
  {\bfseries 56} (2011) 389} [\href{https://arxiv.org/abs/1011.3304}{{\ttfamily
  1011.3304}}].

\bibitem{Shchigolev:2012rp}
V.K.~Shchigolev, \emph{{Cosmic Evolution in Fractional Action Cosmology}},
  \href{https://doi.org/10.5890/DNC.2013.04.002}{\emph{Discontinuity
  Nonlinearity and Complexity} {\bfseries 2} (2013) 115}
  [\href{https://arxiv.org/abs/1208.3454}{{\ttfamily 1208.3454}}].

\bibitem{Shchigolev:2013jq}
V.K.~Shchigolev, \emph{{Fractional Einstein-Hilbert Action Cosmology}},
  \href{https://doi.org/10.1142/S0217732313500569}{\emph{Mod. Phys. Lett. A}
  {\bfseries 28} (2013) 1350056}
  [\href{https://arxiv.org/abs/1301.7198}{{\ttfamily 1301.7198}}].

\bibitem{Shchigolev:2015rei}
V.K.~Shchigolev, \emph{{Testing Fractional Action Cosmology}},
  \href{https://doi.org/10.1140/epjp/i2016-16256-6}{\emph{Eur. Phys. J. Plus}
  {\bfseries 131} (2016) 256}
  [\href{https://arxiv.org/abs/1512.04113}{{\ttfamily 1512.04113}}].

\bibitem{Calcagni:2016ofu}
G.~Calcagni, S.~Kuroyanagi and S.~Tsujikawa, \emph{{Cosmic microwave background
  and inflation in multi-fractional spacetimes}},
  \href{https://doi.org/10.1088/1475-7516/2016/08/039}{\emph{JCAP} {\bfseries
  08} (2016) 039} [\href{https://arxiv.org/abs/1606.08449}{{\ttfamily
  1606.08449}}].

\bibitem{Calcagni:2020ads}
G.~Calcagni and A.~De~Felice, \emph{{Dark energy in multifractional
  spacetimes}}, \href{https://doi.org/10.1103/PhysRevD.102.103529}{\emph{Phys.
  Rev. D} {\bfseries 102} (2020) 103529}
  [\href{https://arxiv.org/abs/2004.02896}{{\ttfamily 2004.02896}}].

\bibitem{Calcagni:2021ipd}
G.~Calcagni, \emph{{Multifractional theories: an updated review}},
  \href{https://doi.org/10.1142/S021773232140006X}{\emph{Mod. Phys. Lett. A}
  {\bfseries 36} (2021) 2140006}
  [\href{https://arxiv.org/abs/2103.06557}{{\ttfamily 2103.06557}}].

\bibitem{Calcagni:2021aap}
G.~Calcagni, \emph{{Classical and quantum gravity with fractional operators}},
  \href{https://doi.org/10.1088/1361-6382/ac1bea}{\emph{Class. Quant. Grav.}
  {\bfseries 38} (2021) 165005}
  [\href{https://arxiv.org/abs/2106.15430}{{\ttfamily 2106.15430}}].

\bibitem{Shchigolev:2021lbm}
V.K.~Shchigolev, \emph{{Fractional-order derivatives in cosmological models of
  accelerated expansion}},
  \href{https://doi.org/10.1142/S0217732321300147}{\emph{Mod. Phys. Lett. A}
  {\bfseries 36} (2021) 2130014}
  [\href{https://arxiv.org/abs/2104.12610}{{\ttfamily 2104.12610}}].

\bibitem{Landim:2021www}
R.G.~Landim, \emph{{Fractional dark energy}},
  \href{https://doi.org/10.1103/PhysRevD.103.083511}{\emph{Phys. Rev. D}
  {\bfseries 103} (2021) 083511}
  [\href{https://arxiv.org/abs/2101.05072}{{\ttfamily 2101.05072}}].

\bibitem{Garcia-Aspeitia:2022uxz}
M.A.~Garc\'\i{}a-Aspeitia, G.~Fernandez-Anaya, A.~Hern\'andez-Almada, G.~Leon
  and J.~Maga\~na, \emph{{Cosmology under the fractional calculus approach}},
  \href{https://doi.org/10.1093/mnras/stac3006}{\emph{Mon. Not. Roy. Astron.
  Soc.} {\bfseries 517} (2022) 4813}
  [\href{https://arxiv.org/abs/2207.00878}{{\ttfamily 2207.00878}}].

\bibitem{Gonzalez:2023who}
E.~Gonz\'alez, G.~Leon and G.~Fernandez-Anaya, \emph{{Exact solutions and
  cosmological constraints in fractional cosmology}},  3, 2023.

\bibitem{Micolta-Riascos:2023mqo}
B.~Micolta-Riascos, A.D.~Millano, G.~Leon, C.~Erices and A.~Paliathanasis,
  \emph{Revisiting fractional cosmology},
  \href{https://doi.org/10.3390/fractalfract7020149}{\emph{Fractal and
  Fractional} {\bfseries 7} (2023) }.

\bibitem{Weinberg:1988cp}
S.~Weinberg, \emph{{The Cosmological Constant Problem}},
  \href{https://doi.org/10.1103/RevModPhys.61.1}{\emph{Rev. Mod. Phys.}
  {\bfseries 61} (1989) 1}.

\bibitem{Zlatev:1998tr}
I.~Zlatev, L.-M.~Wang and P.J.~Steinhardt, \emph{{Quintessence, cosmic
  coincidence, and the cosmological constant}},
  \href{https://doi.org/10.1103/PhysRevLett.82.896}{\emph{Phys. Rev. Lett.}
  {\bfseries 82} (1999) 896}
  [\href{https://arxiv.org/abs/astro-ph/9807002}{{\ttfamily
  astro-ph/9807002}}].

\bibitem{Riess:2021jrx}
A.G.~Riess et~al., \emph{{A Comprehensive Measurement of the Local Value of the
  Hubble Constant with 1 km s$^{-1}$ Mpc$^{-1}$ Uncertainty from the Hubble
  Space Telescope and the SH0ES Team}},
  \href{https://doi.org/10.3847/2041-8213/ac5c5b}{\emph{Astrophys. J. Lett.}
  {\bfseries 934} (2022) L7}
  [\href{https://arxiv.org/abs/2112.04510}{{\ttfamily 2112.04510}}].

\bibitem{Planck:2018vyg}
{\scshape Planck} collaboration, \emph{{Planck 2018 results. VI. Cosmological
  parameters}},
  \href{https://doi.org/10.1051/0004-6361/201833910}{\emph{Astron. Astrophys.}
  {\bfseries 641} (2020) A6}
  [\href{https://arxiv.org/abs/1807.06209}{{\ttfamily 1807.06209}}].

\bibitem{Riess:2019cxk}
A.G.~Riess, S.~Casertano, W.~Yuan, L.M.~Macri and D.~Scolnic, \emph{{Large
  Magellanic Cloud Cepheid Standards Provide a 1\% Foundation for the
  Determination of the Hubble Constant and Stronger Evidence for Physics beyond
  $\Lambda$CDM}},
  \href{https://doi.org/10.3847/1538-4357/ab1422}{\emph{Astrophys. J.}
  {\bfseries 876} (2019) 85}
  [\href{https://arxiv.org/abs/1903.07603}{{\ttfamily 1903.07603}}].

\bibitem{DiValentino:2021izs}
E.~Di~Valentino, O.~Mena, S.~Pan, L.~Visinelli, W.~Yang, A.~Melchiorri et~al.,
  \emph{{In the realm of the Hubble tension\textemdash{}a review of
  solutions}}, \href{https://doi.org/10.1088/1361-6382/ac086d}{\emph{Class.
  Quant. Grav.} {\bfseries 38} (2021) 153001}
  [\href{https://arxiv.org/abs/2103.01183}{{\ttfamily 2103.01183}}].
  

\bibitem{Motta:2021hvl}
V.~Motta, M.~A.~Garc\'\i{}a-Aspeitia, A.~Hern\'andez-Almada, J.~Maga\~na and T.~Verdugo,
 \emph{{Taxonomy of Dark Energy Models}}, 
\href{https://doi.org/10.3390/universe7060163}{\emph{Universe} {\bfseries 7} (2021) no.6, 163}
[\href{https://arxiv.org/abs/2104.04642}{{\ttfamily 2104.04642}}].

\bibitem{Foreman_emcee_2013}
D.~{Foreman-Mackey}, A.~{Conley}, W.~{Meierjurgen Farr}, D.W.~{Hogg},
  D.~{Lang}, P.~{Marshall} et~al., \emph{{emcee: The MCMC Hammer}},
  \href{https://doi.org/10.1086/670067}{\emph{Publ. Astron. Soc. Pac.}
  {\bfseries 125} (2013) 306}
  [\href{https://arxiv.org/abs/arXiv:1202.3665}{{\ttfamily arXiv:1202.3665}}].

\bibitem{Moresco:2016mzx}
M.~Moresco, L.~Pozzetti, A.~Cimatti, R.~Jimenez, C.~Maraston, L.~Verde et~al.,
  \emph{{A 6\% measurement of the Hubble parameter at $z\sim0.45$: direct
  evidence of the epoch of cosmic re-acceleration}},
  \href{https://doi.org/10.1088/1475-7516/2016/05/014}{\emph{JCAP} {\bfseries
  05} (2016) 014} [\href{https://arxiv.org/abs/1601.01701}{{\ttfamily
  1601.01701}}].

\bibitem{Scolnic:2018}
D.M.~Scolnic et~al., \emph{{The Complete Light-curve Sample of
  Spectroscopically Confirmed SNe Ia from Pan-STARRS1 and Cosmological
  Constraints from the Combined Pantheon Sample}},
  \href{https://doi.org/10.3847/1538-4357/aab9bb}{\emph{Astrophys. J.}
  {\bfseries 859} (2018) 101}
  [\href{https://arxiv.org/abs/1710.00845}{{\ttfamily 1710.00845}}].

\bibitem{Scolnic_Complete_2018}
D.M.~{Scolnic}, D.O.~{Jones}, A.~{Rest}, Y.C.~{Pan}, R.~{Chornock},
  R.J.~{Foley} et~al., \emph{{The Complete Light-curve Sample of
  Spectroscopically Confirmed SNe Ia from Pan-STARRS1 and Cosmological
  Constraints from the Combined Pantheon Sample}},
  \href{https://doi.org/10.3847/1538-4357/aab9bb}{\emph{Astrophys. J.}
  {\bfseries 859} (2018) 101}
  [\href{https://arxiv.org/abs/arXiv:1710.00845}{{\ttfamily
  arXiv:1710.00845}}].

\bibitem{Magana_Cardassian_2018}
J.~{Maga{\~n}a}, M.H.~{Amante}, M.A.~{Garcia-Aspeitia} and V.~{Motta},
  \emph{{The Cardassian expansion revisited: constraints from updated Hubble
  parameter measurements and type Ia supernova data}},
  \href{https://doi.org/10.1093/mnras/sty260}{\emph{Mon. Not. Roy. Astron.
  Soc.} {\bfseries 476} (2018) 1036}
  [\href{https://arxiv.org/abs/arXiv:1706.09848}{{\ttfamily
  arXiv:1706.09848}}].

\bibitem{Zhao:2017cud}
G.-B.~Zhao et~al., \emph{{Dynamical dark energy in light of the latest
  observations}}, \href{https://doi.org/10.1038/s41550-017-0216-z}{\emph{Nature
  Astron.} {\bfseries 1} (2017) 627}
  [\href{https://arxiv.org/abs/1701.08165}{{\ttfamily 1701.08165}}].

\bibitem{H0diagnostic:2021}
C.~Krishnan, E.~Ó~Colgáin, M.M.~Sheikh-Jabbari and T.~Yang, \emph{Running
  hubble tension and a h0 diagnostic},
  \href{https://doi.org/10.1103/physrevd.103.103509}{\emph{Phys. Rev. D}
  {\bfseries 103} (2021) }.

\bibitem{Valcin:2021}
D.~Valcin, R.~Jimenez, L.~Verde, J.L.~Bernal and B.D.~Wandelt, \emph{The age of
  the universe with globular clusters: reducing systematic uncertainties},
  \href{https://doi.org/10.1088/1475-7516/2021/08/017}{\emph{Journal of
  Cosmology and Astroparticle Physics} {\bfseries 2021} (2021) 017}.

\end{thebibliography}

\end{document}